\documentclass[printer]{tADP2e}
\usepackage{graphicx}
\begin{document}
\markboth{Wet granular materials}{Namiko Mitarai and Franco Nori}
\title{Wet granular materials}
\author{Namiko Mitarai$^{\dagger*}$
\footnote{*Email: namiko@stat.phys.kyushu-u.ac.jp} 
and Franco 
Nori$^{\ddagger,\dagger\dagger}$ 
\newline\centerline{
$^\dagger$ Department of 
Physics, Kyushu University 33, Fukuoka 812-8581, Japan.}
\newline 
$^\ddagger$Frontier Research System,
The Institute of Physical and Chemical Research (RIKEN),
Hirosawa 2-1, Wako-shi, Saitama 351-0198, Japan.
\newline 
$^{\dagger\dagger}$Center for Theoretical Physics, Physics Department, 
Center for the Study of Complex Systems,
The University of Michigan, Ann Arbor, Michigan 48109-1040,
USA}

\maketitle
\begin{abstract} 
Most studies on granular physics have focused on 
dry granular media, with no liquids between the grains.
However, in geology and many real world applications
(e.g., food processing, pharmaceuticals, ceramics,
civil engineering, constructions, and many industrial applications),
liquid is present between the grains. 
This produces inter-grain cohesion
and drastically modifies the mechanical properties 
of the granular media (e.g., the
surface angle can be larger than 90 degrees).
Here we present a review of the mechanical 
properties of wet granular media, with
particular emphasis on the effect of cohesion.
We also list several open problems that 
might motivate future studies in this exciting but 
mostly unexplored field.

\end{abstract}                                                                 
\section{Introduction}
\label{sec:intro}
\subsection{Granular physics and wet granular media}
Granular materials are collections of macroscopic particles,
like glass beads or sand,
which are visible by the naked eye.
Because of the macroscopic size of the particles,
thermal noise plays no role in particle motion,
and particle-particle interactions are dissipative.
Therefore, continuous energy input by external forces
(gravity, vibrations, etc.) are necessary in order to keep them in motion.
Particles may stay at rest like a solid, flow like a liquid,
or behave like a gas,
depending on the rate of energy input.
However, the external force is often not enough for
the particles to explore their phase space,
which makes them 
quite different from conventional molecular systems.

The scientific study of granular media
has a long history,
mainly in the engineering field,
and many physicists have joined the granular research community
over the past few decades
(for reviews and books, see, e.g.,
\cite{Duran97,Jaeger96,Nedderman,deGennesGranular,
PandG97,BagnoldBook,Jaeger92,Jaeger96-PT,Sholtz97,Nori97,
Chaos9,Edwards02}).
Most studies on granular media, especially in the physics field,
have focused on {\it dry granular materials},
where the effects of 
interstitial fluids are negligible 
for the particle dynamics.
For dry granular media,
the dominant interactions are inelastic collisions and friction,
which are short-range and {\it non-cohesive}.
Even in this idealised situation, dry granular media
show unique and striking behaviours, which have attracted
the attention of many scientists for centuries.

In the real world, however, 
we often see {\it wet granular materials},
such as beach sand.
Dry and wet granular materials have many aspects in common, 
but there is one big difference:
Wet granular materials are {\it cohesive}
due to surface tension.

In the past, many research groups that studied 
granular physics have been struggling to
minimise humidity and avoid inter-granular
cohesive forces (e.g., \cite{Bretz92}).
Indeed, some experiments were performed in 
vacuum  chambers.
Humidity and fluids in general were
seen as a nuisance to be avoided at all costs.
However, many important real life applications
involve mechanical properties of wet granular media.
Examples include rain-induced landslides,
pharmaceuticals,
food processing,
mining and construction industries.
Thus, it is important to study the mechanical 
response of granular matter with various degrees of 
wetness or liquid content.

In this review, we show how the cohesion 
induced by the liquid changes the mechanical properties of
granular materials.
We mainly consider static or quasistatic situations,
where the cohesion dominates over other effects of the liquid,
such as lubrication and viscosity.
Some phenomena in this field
which are not well-understood are presented below as ``open problems''.
Most references listed at the end focus on experimental results.
Theoretical approaches can be found in the
reviews cited below and references therein.

\subsection{What is different from dry granular media}
\begin{table*}
\begin{center}
\caption{
Comparison of physical properties between dry 
and wet granular matter.}
\label{Table:DryWet}
\begin{tabular}[t]{|p{3.cm}||p{3.5cm}|p{4.5cm}|}
\hline
PROPERTY&DRY & WET \\
\hline
\hline
Cohesion&Negligible&Important\\
\hline
Surface angle 
& Finite\newline
Around $35^\circ$ for sand&
Finite: Larger than the dry case\newline
Can be as large as $90^\circ$
, or even larger\\
\hline
Tensile strength&
Negligible&
Finite \\
\hline
Yield shear stress & Finite \newline 
Zero at zero normal stress
 & 
Finite: Can be larger than the dry case\newline
Non-zero at zero normal stress\\
\hline
Hysteresis&Yes&Yes: Enhanced\\
\hline
Configurational phase space 
for packing
& Finite 
 & 
Finite: Can be larger than the dry case
\\
\hline
\end{tabular}
\end{center}
\end{table*}
Studies of wet granular media
have been made in 
many industrial applications.
The mechanical properties of wet
granular media are also 
extremely important
in geology and civil engineering.
For example, 
let us consider the huge and expensive civil works project of
the construction of the Kansai international airport, 
on a man-made island near Osaka.
Because of the weight of the 
180 million cubic meters of landfill and facilities, 
the seabed composed of clay is compressed, and 
it is inevitable for the airport to sink by some amount; 
The airport had sunk by 11.7 meters on average at the end of 2000,
and the settlement is still carefully monitored \cite{KIX}. 
This and other examples from geology 
and civil engineering stress the need
to better understand
the mechanical properties of wet granular 
assemblies, both small and large.

The biggest effect that the liquid in granular media induces
is the cohesion between grains.
Even humidity in the air may
result in a tiny liquid bridge at a contact point,
which introduces cohesion.
The cohesion occurs in wet granular material
unless the system becomes over-wet, i.e., 
the granular medium is completely immersed in a liquid.
In this short review, we focus on the effect of this cohesion;
the system that we are considering is partially wet
granular material, which is a mixture of solid grains, liquid, and air.

In addition to cohesion, there are many effects 
induced by the presence of the liquid. 
One of them is the lubrication of
solid-solid friction \cite{Tribology,LubricationE,FrictionBook}. 
In addition, the liquid viscosity may induce
a velocity-dependent behaviour 
and additional dissipation.
These effects are often seen 
in underwater experiments 
(e.g., \cite{Geminard99,Medved01}).
The time scale of liquid motion
(how the liquid moves or flows through granular media)
also affects the dynamics.
All of these effects, of course, play
important roles in the properties of wet granular media.
However, these are more or less velocity-dependent
phenomena, and in the static or quasistatic regime,
the cohesion often plays the most important role,
providing a significant qualitative difference from dry granular media.

\begin{figure}
\begin{center}
\epsfxsize=0.4\textwidth
\epsfbox{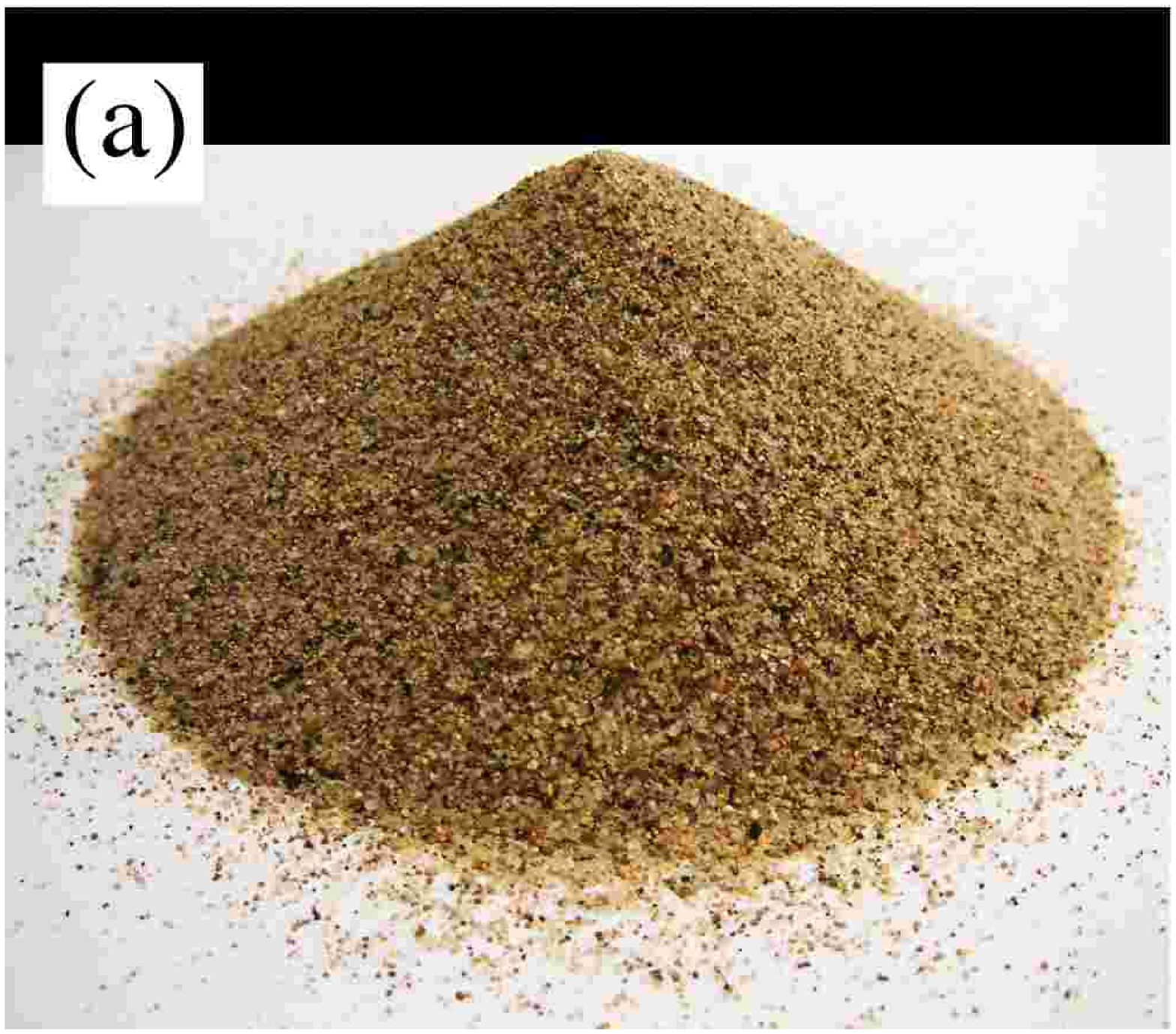}
\epsfxsize=0.41\textwidth
\epsfbox{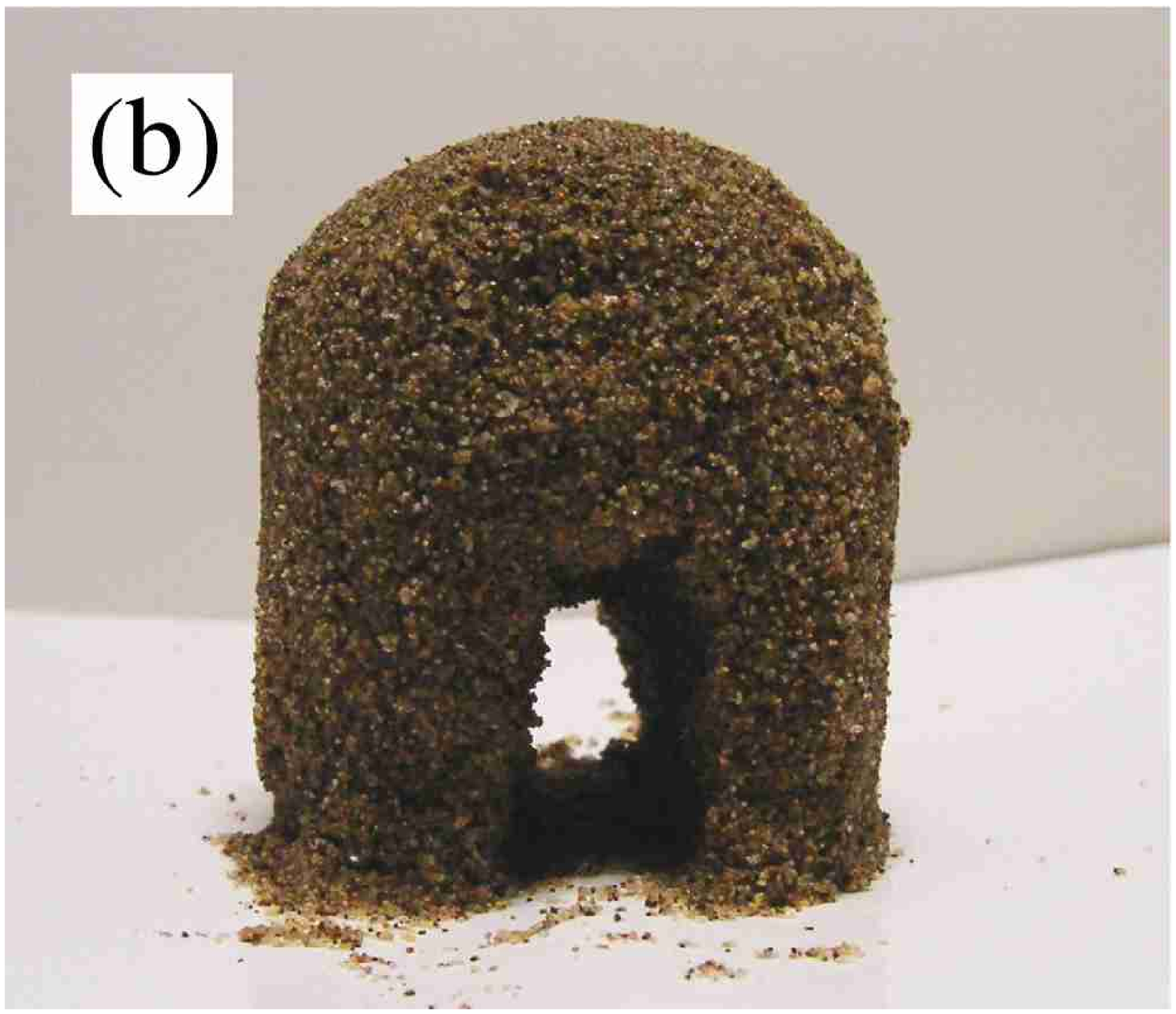}
\end{center}
\caption{
(a) Dry sandpile with a well-defined surface angle. 
(b) Wet sandpile with a tunnel.
}
\label{Fig:Sandpile}
\end{figure}
\begin{figure}
\begin{center}
\epsfxsize=0.5\textwidth
\epsfbox{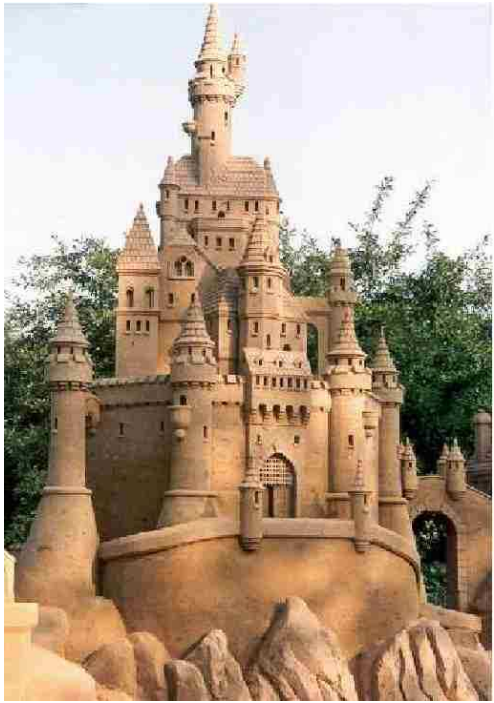}
\end{center}
\caption{
Sandcastle made of wet sand 
(Copyright Sandscapes 2005).
}
\label{Fig:Sandcastle}
\end{figure}

The simplest situation where we see the 
effect of cohesion in wet granular media
would be the sandpiles that children make
in a sandbox.
Let us first consider a sandpile made of dry grains 
and afterwards wet grains.

When we make a sandpile using dry sand
(Fig.~\ref{Fig:Sandpile}(a)),
the surface of the pile is smooth, 
and makes a finite, well-defined angle, which is about $35^\circ$.
A slightly denser pile might be made by tapping the surface of the pile,
but it does not change much the shape of the pile.
Even if we try to make a sharper pile by pouring 
more dry sand, grains flow down along
the pile surface and the resulting angle is always around
the same value.
It is easy to remove part of the sand from the pile, 
but we cannot make a tunnel through dry sandpile
because the wall around the hole would collapse and the angle 
of the surface cannot be larger than the critical angle.

Let us add some water to the sand,
and make another sandpile (Fig.~\ref{Fig:Sandpile}(b)).
We can try to find the optimal amount of 
water to produce a certain shape.
Small amounts of water do not 
allow the creation of many shapes,
while too much water results in muddy water
that cannot keep a shape. 
With the proper amount of water,
we can make a sandpile with 
the surface angle larger than the dry case;
Indeed, the angle can be as large as 
$90^\circ$, or even larger,
allowing the construction of elaborate and stunning sandcastles
(see, e.g., Fig.~\ref{Fig:Sandcastle}).
This wet pile can be made denser and stronger 
by tapping the surface.
Now we can make a tunnel through the pile,
if we are sufficiently careful. 
If we try to make a hole that is too large, 
the pile would break down
forming some rugged surfaces.

Therefore, in a sandbox, we already learnt
important properties of dry granular media:
a finite angle of repose, small hysteresis in packing,
and small strength against loading.
We also learnt how drastically these properties 
are changed or enhanced by adding a liquid,
resulting in: 
a much larger angle of repose,
stronger hysteresis in packing,
and enhanced strength against loading.
These comparisons are summarised in Table \ref{Table:DryWet}.
In the following sections, 
we will see how these behaviours are studied by scientists.

\section{Wet granular media: Grains with liquid and air}
\subsection{Cohesion between two spheres}
\subsubsection{Meniscus and suction}
Cohesion in wet granular media arises from surface tension 
and capillary effects of the liquid.
Consider a meniscus between 
air with pressure $P_a$ and liquid with pressure $P_l$.
The pressure difference $\Delta P$ between liquid and air 
with a meniscus of curvature radii $r_1$ and $r_2$ 
is given by the Young-Laplace equation as
\begin{equation}
\Delta P=P_a-P_l=\gamma\left[\frac{1}{r_1}+\frac{1}{r_2}\right],
\label{eq:DeltaP}
\end{equation}
where $\gamma$ is the surface tension between air and the liquid,
and the curvature is positive 
when the meniscus is drawn back into the liquid phase
(e.g., \cite{FluidMechanics}).
When the curvature is positive,
$\Delta P$ is positive, and
it is often called the suction.

The capillary length 
\begin{equation}
a=\sqrt{\frac{2 \gamma}{\rho_l g}}
\end{equation}
gives the length scale that compares
the capillary force and gravity, where $g$ is
the gravitational acceleration and $\rho_l$
is the mass density of the liquid;
$a$ is around $3.9$ mm for water at room temperature.
The capillary force becomes dominant when 
the relevant length scales are much smaller than $a$. 
Hereafter we consider the situation where 
the capillary force is dominant.
In general, 
one needs to consider the capillary length
when interpreting experimental results,
because most experiments are conducted under gravity.

\subsubsection{Liquid bridge between two spheres}
\label{CohesionTwoParticles}
\begin{figure}
\begin{center}
\includegraphics[width=0.5\textwidth]
{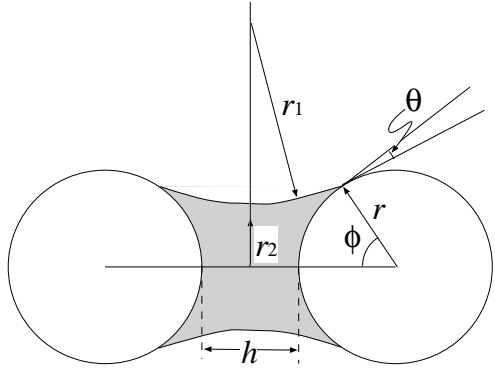}
\end{center}
\caption{
Schematic diagram of a liquid bridge
between identical spheres.
}
\label{Fig:LiquidBridge}
\end{figure}

Now let us see how the liquid induces cohesion in granular media
by considering a liquid bridge between two identical spheres 
as shown in Fig.~\ref{Fig:LiquidBridge}.
The attracting force between spheres due to the liquid menisci
is given by the sum of the surface tension
and the suction; 
when we estimate the force at the neck of the bridge,
it is given by 
\begin{equation}
F_{\rm bridge}=2\pi r_2\gamma+\pi r_2^2\Delta P,
\label{eq:bridge0}
\end{equation}
with
\begin{equation}
\Delta P=\gamma\left[\frac{1}{r_1}-\frac{1}{r_2}\right].
\end{equation}
The detailed experimental analysis of the cohesion force
due to the liquid bridge between two spheres
is found in \cite{Willett00}.
The effect of roughness on the cohesive force
is discussed in detail in \cite{Herminghaus05}.

In real partially wet granular media, 
the picture of a liquid bridge 
formed between completely spherical particles
is often not sufficient to describe the interaction.
However, the presence of a liquid,
which tries to minimise its surface area, generally
results in suction and a cohesive force between particles.

\subsection{Wet granular media with various liquid content}
\subsubsection{Four states of liquid content:
pendular, funicular, capillary, and slurry state}
\begin{sidewaystable}
\caption{
Granular media with various
amounts of liquid \cite{Newitt58,GranulationReview}.
In the schematic diagrams 
in the third column,
the filled circles 
represent the grains and
the grey regions represent the interstitial liquid.
}
\label{Table:Regimes}
\begin{tabular}{|p{1in}||p{.8in}|p{1.3in}|p{3.in}|}
\hline
\begin{minipage}{1in}
{\bf Liquid content}
\end{minipage}
&
\begin{minipage}{.8in}
{\bf State}
\end{minipage}
&
\begin{minipage}{1.3in}
{\bf Schematic diagram}
\end{minipage}
&
\begin{minipage}{3.in}
{\bf Physical description}
\end{minipage}\\
\hline
\hline
\begin{minipage}{1in}
No
\end{minipage}
&
\begin{minipage}{.8in}
Dry
\end{minipage}
&
\begin{minipage}{1.3in}
\includegraphics[width=1in]{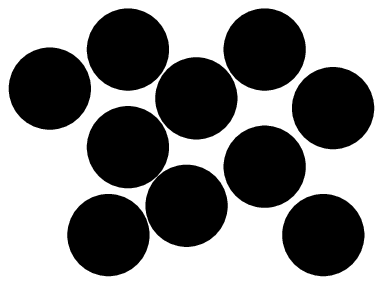}
\end{minipage}
&
\begin{minipage}{3in}
\begin{flushleft}
Cohesion between grains is negligible.
\end{flushleft}
\end{minipage}
\\
\hline
\begin{minipage}{1in}
Small 
\end{minipage}
&
\begin{minipage}{.8in}
Pendular 
\end{minipage}
&
\begin{minipage}{1.3in}
 \includegraphics[width=1in]{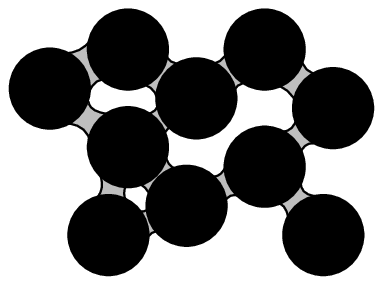}
\end{minipage}
&
\begin{minipage}{3in}
\begin{flushleft}
Liquid bridges are formed at the contact points of grains.
Cohesive forces act through the liquid bridges.
\end{flushleft}
\end{minipage} 
\\
\hline
\begin{minipage}{1in}
Middle
\end{minipage}
& 
\begin{minipage}{.8in}
Funicular
\end{minipage}&
\begin{minipage}{1.3in}
\includegraphics[width=1in]{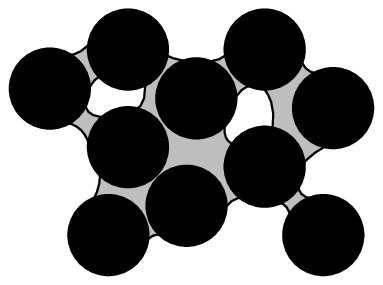}
\end{minipage}
&
\begin{minipage}{3in}
\begin{flushleft}
Liquid bridges around the contact points
and liquid-filled pores coexist.
Both give rise to cohesion between particles.
\end{flushleft}
\end{minipage}
\\
\hline
\begin{minipage}{1in}
Almost saturated
\end{minipage}
&
\begin{minipage}{.8in}
Capillary
\end{minipage}
&
\begin{minipage}{1.3in}
\includegraphics[width=1in]{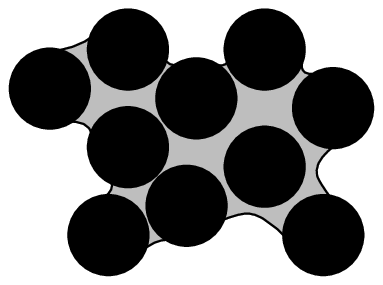}
\end{minipage}
&
\begin{minipage}{3in}
\begin{flushleft}
Almost all the pores are filled with the liquid,
but the liquid surface forms menisci and the 
liquid pressure is lower than the air pressure.
This suction results in a cohesive interaction between particles.
\end{flushleft}
\end{minipage}
\\
\hline
\begin{minipage}{1in}
More 
\end{minipage}& 
\begin{minipage}{.8in}
Slurry 
\end{minipage}
&
\begin{minipage}{1.3in}
\includegraphics[width=1in]{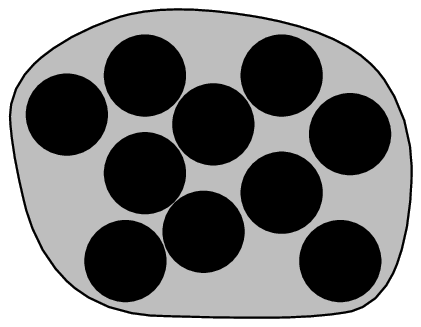}
\end{minipage}
&
\begin{minipage}{3in}
\begin{flushleft}
The liquid pressure is equal to, or higher than, the air
pressure. No cohesive interaction appears between particles.
\end{flushleft}
\end{minipage}
\\
\hline
\end{tabular}
\end{sidewaystable}
\label{phases}

It is well known that cohesion in wet granular materials 
depends on the amount of liquid in the system.
The following four regimes of liquid content
have been distinguished 
in wet granular media \cite{Newitt58,GranulationReview}:
\begin{itemize}
\item Pendular state:  
Particles are held together by liquid bridges at their contact points.

\item Funicular state: 
Some pores are fully saturated by liquid,
but there still remain voids filled with air.

\item Capillary state: 
All voids between particles are filled with liquid, 
but the surface liquid is drawn back into the pores under capillary
action.

\item Slurry state: 
Particles are fully immersed in liquid and the surface of liquid
is convex, i.e., no capillary action at the surface.
\end{itemize}
These four regimes are schematically shown in 
Table \ref{Table:Regimes}.

Cohesion arises in the pendular, funicular,
and capillary states. 
Thus, we  consider these three states in this review.
A priori, the mechanical properties of these three states 
would be expected to be qualitatively different.
In the case of the pendular state,
the cohesive force between a pair of grains
acts through a liquid bridge;
while in the capillary state
the interface between 
the liquid and the air
is pressed due to the suction and 
that pressure keeps together all the grains in the liquid phase.
Both the two-body cohesion due to liquid bridges 
and the suction at the liquid-air interfaces
play important roles in the funicular state.

It is very difficult to directly observe the liquid 
distribution in three-dimensional granular assemblies,
though the liquid distribution should be readily observable
for two-dimensional wet granular aggregates
confined by Plexiglas plates (cf. \cite{Rubio89}). 
As we will see later, 
liquid bridges in three dimensions for rather small 
liquid content have recently been visualised by using index matching 
techniques \cite{Mason99,Fournier05,Herminghaus05},
but it seems to be difficult to extend the method for 
much larger liquid content.
The detailed liquid distribution 
in the each liquid content regimes described in Table
\ref{Table:Regimes} still remains as one of many ``open problems'' 
that we list below as areas that have not been 
sufficiently well studied so far.

Conventionally, these liquid content regimes 
have been distinguished by
measuring the relation between the liquid content $S$
and the suction $\Delta P$ \cite{Newitt58} 
as we will see below,
where $S$ is 
the ratio of the volume of liquid $V_l$ in the system 
to the volume of the voids $V_v$ in the granular media
(when the total volume of the system is
$V_t$ and the volume occupied by grains is $V_s$,
then $V_v=V_t-V_s$ and $V_v=V_l+V_a$
where $V_a$ is the air volume in the system).
Some authors define $S$ as the 
percentage of the liquid content 
(i.e., multiplying the 
ratio of $S=V_l/V_v$ by 100).
Here, either case, ratio or percentage, is clear from the text.
\subsubsection{Liquid content and suction}
\paragraph{Measurement of suction in granular media}
\label{suctionmeasure}
\begin{table*}
\caption{List of methods to measure
the suction and practical suction range for each measurement.
Methods based on the osmotic pressure difference (e.g., due to 
gradients of solute concentration) are not listed below.
Adapted from \cite{UnsaturatedBook}.
}
\label{Table:UnsaturatedBookTable10-1}
\begin{center}
\begin{tabular}{|p{2.15in}|p{1.5in}|p{0.85in}|}
\hline
{\bf Technique} or {\bf Sensor} & {\bf Suction Range} (kPa)
& References\\
\hline
\hline
Tensiometers & 0-100& 
\cite{Cassel86,Stannard92}\\
\hline
Axis translation techniques &
0-1500& 
\cite{Hilf56,Bocking80}\\
\hline
Electrical or thermal conductivity sensors&
0-400&
\cite{Phene71a,Phene71b,Fredlund89}\\
\hline
Contact filter paper method&
Entire range
& 
\cite{Houston94}\\
\hline
\end{tabular}
\end{center}
\end{table*}

There are several ways to measure the suction
$\Delta P$ in granular media,
and the appropriate method should be chosen
regarding the measurement range of $\Delta P$.
Table \ref{Table:UnsaturatedBookTable10-1} 
shows the list of methods and the range of measurements
\cite{UnsaturatedBook}
commonly used in soil mechanics
(where the mixture of soil grains and water
is considered as a specimen).\footnote{
When solutes are dissolved in water in soil,
the osmotic effects produce the chemical potential 
difference from free water,
where free water is the water that contains no dissolved solute,
has no interactions with other phases that produce 
curvature to the air-water interface, and has
no external force other than gravity \cite{UnsaturatedBook}.
In the soil mechanics, this chemical potential difference
is sometimes measured in the unit of pressure,
and it is referred to as osmotic suction.
The pressure difference between the 
air and liquid (water) in soil $\Delta P=P_a-P_l$, which is 
due to capillary effect and 
short-range adsorption of water to grain surfaces,
is referred to as matric suction.
In this paper, we focus on the latter,
and $\Delta P$ is simply referred to as suction.
}
It is beyond the scope of this brief review
to describe all the experimental methods used,
but theses are described in the references.
As examples, below we describe two
often-used methods, 
called the axis-translation techniques and 
tensiometers,
which make use of the capillary pressure
in a porous ceramic disk to measure and to control the suction
\cite{UnsaturatedBook}.

\begin{figure}
\begin{center}
\includegraphics[width=1.\textwidth]{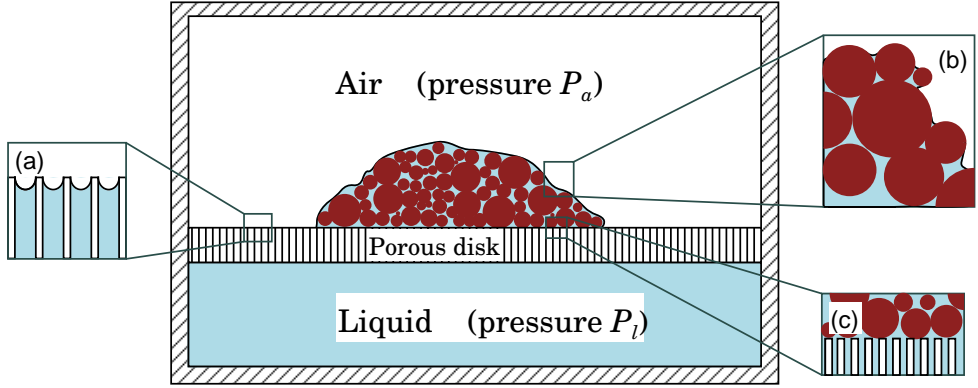}
\end{center}
\caption{
Schematic diagram describing the 
tensiometers and the axis-translation techniques.
The filled circles represent grains
and the light blue region represents a liquid.
The container is separated by a High-Air-Entry (HAE) 
porous disk that has many microscopic pores.
The upper part is filled with air with pressure $P_a$,
and the lower part is filled with liquid with pressure $P_l$.
The pores in the porous disk are filled with liquid,
and the menisci at the top surface of the pores in the porous disk
(a magnified view is shown in inset (a))
make it possible to keep $P_l<P_a$.
A wet granular assembly is placed on the porous disk 
(a magnified view is shown in inset (b)),
and liquid in the wet granular media
is in good contact with the liquid in the porous disk 
(a magnified view is shown in inset (c));
the liquid pressure around the wet grains
equals the pressure $P_l$
of the bulk liquid in the bottom of the container.
Adapted from 
\cite{UnsaturatedBook}. 
}
\label{Fig:UnsaturatedBookFig5-19}
\end{figure}

Relatively low suction can be measured or controlled
by using tensiometers and the axis-translation technique
\cite{UnsaturatedBook}.
A schematic diagram to describe these methods 
is given in Fig.~\ref{Fig:UnsaturatedBookFig5-19};
These methods make use of the properties of 
what is called a ``high-air-entry (HAE) material'',
which is a material with many microscopic pores,
such as a porous ceramic. 
In Fig.~\ref{Fig:UnsaturatedBookFig5-19},
a container is separated in two parts by a HAE disk;
the bottom part of the box is filled with a liquid 
with pressure $P_l$, while
the upper part is filled with air with pressure $P_a$.
A magnified view of the HAE disk is given in the inset (a),
where the pores of the HAE material are saturated with liquid.
There, the surface tension at the liquid-air interfaces
formed in the pores allow a finite pressure difference between 
the liquid and air.
Considering Eq.~(\ref{eq:DeltaP})
and the magnified view (a) in Fig.~\ref{Fig:UnsaturatedBookFig5-19},
the largest possible pressure difference 
between the air and liquid in the HAE material is roughly given by 
\begin{equation}
\Delta P_{\max}=\frac{2\gamma}{R_{\rm max}}.
\label{Eq:deltapmax}
\end{equation}
Here, as a first approximation, 
we assume that the pores are tube-shaped with 
maximum radius  $R_{\rm max}$;
for smaller pore size, $\Delta P_{\max}$ would be larger.
The liquid pressure $P_l$ in the lower part of the container
and the air pressure $P_a$ in the upper part 
can be separately controlled 
without allowing the air to enter the lower part of the container,
as far as $\ (P_a-P_l)<\Delta P_{\max}\ $ is satisfied.

Now, let us place a wet granular material on the HAE disk
inside the container (Fig.~\ref{Fig:UnsaturatedBookFig5-19}, inset (b)) 
\cite{UnsaturatedBook}.
When the equilibrium is reached after
a long enough waiting time, the 
liquid pressure in the wet granular medium should be equal to
the liquid pressure $P_l$ in the lower part of the container,
if the liquid phase in the granular medium is in good contact with
the liquid in the pores of the HAE disk
(Fig.~\ref{Fig:UnsaturatedBookFig5-19}, inset (c)).
The air pressure $P_a$ and the liquid pressure $P_l$
in the upper and lower parts of the container, respectively,
can be easily measured, or even controlled,
and the suction $\Delta P$ in the wet granular material 
is given by $\Delta P=P_a-P_l$
\cite{UnsaturatedBook}.

The method to control the suction $\Delta P$
by controlling the liquid and air pressures separately 
is called the axis-translation technique \cite{UnsaturatedBook}.
Initially a saturated granular material is placed on 
a HAE disk that separates the liquid and the air, as 
schematically shown in 
Fig.~\ref{Fig:UnsaturatedBookFig5-19},
and the liquid in the granular assembly is drained
until the system reaches the equilibrium, where 
the liquid pressure in the granular medium becomes equal to 
that of the bulk liquid in the lower box.
The tool to measure the suction by using the HAE disk
is called a tensiometer, which is 
composed of a cup made of a HAE ceramic and
a sensor to measure the liquid pressure, 
connected by a tube filled with liquid (water) \cite{UnsaturatedBook}.
The HAE ceramic cup 
is placed in the wet granular assembly,
and the pore pressure is measured by a direct exchange of liquid between 
the sensor and the wet granular medium.
Note that, in using either 
the axis-translation technique or
a tensiometer,
the maximum possible value of the suction is limited by $\Delta P_{\max}$;
other methods used to measure larger pressure differences
are listed in Table \ref{Table:UnsaturatedBookTable10-1} 
and Ref. \cite{UnsaturatedBook}.

\paragraph{Relation between the liquid content and suction}
\begin{figure}
\begin{center}
\includegraphics[width=0.5\textwidth]{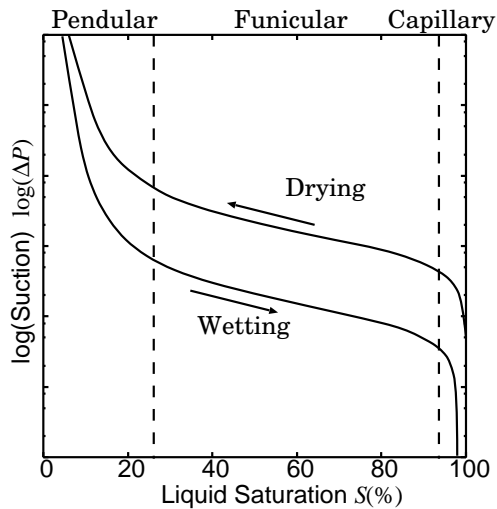}
\end{center}
\caption{
A schematic plot of 
the suction $\Delta P$ versus the liquid content $S$. 
The suction is shown in a log-scale.
The upper curve represents the drying process,
and the lower curve represents the wetting process.
It changes the slope significantly near the phase boundaries between 
the pendular, funicular, and capillary states.
}
\label{Fig:SoilWater}
\end{figure}

The schematic description of the suction $\Delta P$
versus the liquid content $S$ 
is shown in Fig.~\ref{Fig:SoilWater}
\cite{UnsaturatedBook,Newitt58}.
Two lines are shown because there is a hysteresis in the wetting
and drying processes.
In both lines, the slope
for low and large liquid content is much larger 
than that for intermediate liquid content.
This change of slope reflects the
liquid distribution in granular media.

For simplicity, let us consider the drying process
(the drainage of liquid)
from completely saturated granular media. 
At first the material is in the slurry state, 
and the liquid pressure 
is the same or larger than the air pressure.
After an appropriate amount of liquid is drained,
the media is in the capillary state, and
the liquid pressure becomes lower than the air pressure,
i.e., the suction $\Delta P$ becomes positive.
When we further decrease the liquid content $S$ 
in the capillary state, 
$\Delta P$ increases very rapidly upon decreasing $S$; 
$\Delta P$ is determined by the curvature
of the liquid surface,
and the change of the surface curvature 
requires little change of liquid content $S$ because
bulk liquid remains in the system.
When $\Delta P$ exceeds a critical value,
which is mainly determined by the surface tension and
the geometry of granular media around the surface liquid,
air starts to enter the bulk of the system and the funicular state
is obtained.
The suction still increases as the liquid content decreases,
but its rate of increase becomes much smaller than that in 
the capillary state,
because the liquid in the pores in the bulk needs to be removed.
At some point, most of the voids will be
filled with air, and the rest of the liquid is
found around the contact points between neighbouring grains 
and in the thin liquid film that covers the grains' surfaces.
Namely, the system is in the pendular state.
In this pendular state, a small change of the liquid 
content again results in a relatively large change of 
the suction, because the total amount of the liquid 
is much smaller than the void volume $V_v$ and
the change of curvature at each liquid bridge requires 
only a small change of the liquid content $S$.

As shown in Fig.~\ref{Fig:SoilWater},
for a given value of the suction $\Delta P$,
the liquid content $S$ is smaller for the wetting process
of initially dry granular media than that of the drying process
of initially saturated grains.
This is partly because of the hysteresis in
the wetting and drying processes at solid surfaces
\cite{deGennesWet,Herminghaus05,QuereBook},
and partly because the variation of the pore size
in the granular medium
\cite{UnsaturatedBook}.
The flow of liquid in porous media
(including granular media)
is an important and interesting problem;
indeed, this problem has developed into 
a separate research area.
For a recent review on this problem, in terms of statistical physics, 
see 
\cite{ImbibitionReview}.

\section{Mechanical properties}
In this section, we summarise what is known about the 
mechanical properties of wet granular media.
As mentioned above, here we focus on granular cohesion
in the static and quasistatic regimes
for varying liquid content.
To also have some idea about the dynamics,
dynamical experiments are briefly presented 
in the final part of this section.
\subsection{
Granular cohesion in the static and quasistatic regimes}
\subsubsection{Compaction of wet granular media}
\label{compaction}
The configuration of grains in a 
container shows history-dependence. 
In the case of dry granular media, 
the material just poured in a container
is compacted when it is tapped.
The dynamics of this compaction process is slow,
which is known to show logarithmic dependence on the number 
of taps (e.g., 
\cite{Knight95,Nowak98,Nicodemi97}).

In the case of wet granular media, 
very low-density configurations can be realised
even when the grains are spherical.
Due to the cohesion, a configuration with many large openings 
or gaps surrounded by chains of particles can be stabilised,
which may not be stable for dry spherical grains 
\cite{Xu04}.
The compaction via a ram
of this low-density configuration
by a quasistatic compression has been examined
\cite{Gioia02,Kong00}.
With increasing pressure on the system,
the relatively large openings between particles
disappear;
a grain in a chain snaps in away from the chain, 
and the chain structure changes.
After the openings disappear, a high-density region is
formed, and the density difference between 
the initial low-density and 
newly-formed high-density regions becomes clearly visible.
One can see that the domain wall 
between the high-density and low-density regions
travels through the system like a wave.

The realisability of very low-density structures
indicates that the hysteresis of configurations
in wet granular media can be stronger 
than that in dry granular media.
In other words, the many 
huge gaps or openings 
that are possible in wet granular media, 
make the configuration phase space available to the grains
much larger than the dry granular case 
(see also Table \ref{Table:DryWet}).

\subsubsection{Angle of repose for small amounts of liquid}
\label{AngleofRepose}
\paragraph{Experiments}
The angle of repose $\theta_r$ and 
the critical angle $\theta_c$ are measurable quantities 
that are sensitive to the cohesion.
Here, the critical angle $\theta_c$ is the surface angle just before an 
avalanche occurs, while the angle of repose $\theta_r$ is the surface angle 
after the avalanche, and is typically slightly smaller than $\theta_c$.
When the amount of liquid is small,
the cohesion is also small, but the cohesion is still 
detectable through $\theta_r$ and $\theta_c$
of either a granular pile or
in a drum that rotates very slowly
\cite{Bocquet98,Hornbaker97,Albert97,Fraysse99,Halsey98,Nase01}.

\begin{figure}
\begin{center}
\includegraphics[width=0.5\textwidth]{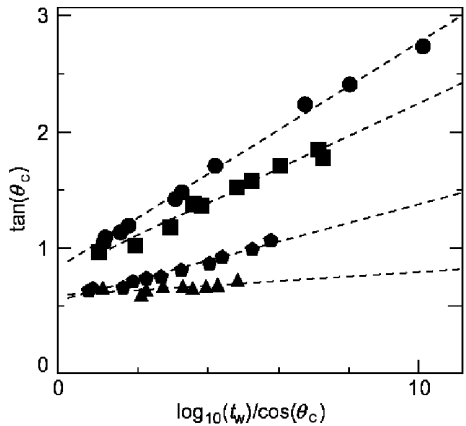}
\end{center}
\caption{
The critical angle $\theta_c$ 
of glass beads in humid air
as a function of the waiting time $t_w$
in a slowly rotating drum,
shown as $\tan\theta_c$ versus $(\log_{10}t_w/\cos\theta_c)$.
The time is in units of seconds,
and humidities are  15\% (triangles),
27\% (pentagons), 36.1\% (squares), and 45.5\% (circles).
As time passes,
more and more liquid condenses 
at the contact points between particles,
increasing the cohesion between them.
Therefore, the critical angle
increases with time.
From 
\cite{Bocquet98} .
}
\label{Fig:Bocquet98Fig2a}
\end{figure}

Even the humidity in the air is enough to
induce cohesion for initially dry granular material
\cite{Bocquet98,Restagno02,Fraysse99,Ovarlez01,DAnna00,
Danna00-2,Forsyth02,Crassous99}.
Bocquet {\it et al.} \cite{Bocquet98}
measured the critical angle $\theta_c$
of glass beads in humid air, and found that
$\theta_c$ depends on the waiting time $t_w$,
namely, how long they waited after preparing the sample
until measurement. 
They reported the relation 
$(\tan\theta_c -\tan\theta_0)\propto \log(t_w)/\cos\theta_c$,
where $\theta_0$ is a constant
(Fig.~\ref{Fig:Bocquet98Fig2a}),
and also found that the $t_w$-dependence 
of $\theta_c$ varies with humidity.

This time dependence 
of the critical angle $\theta_c$ on the waiting time $t_w$
has been analysed based on
the capillary condensation, 
which results from the fact that
the equilibrium vapour pressure
in a narrow space where
non-zero suction $\Delta P$ is allowed
is lower than that in the bulk
\cite{Israelachvili}.
\footnote{ 
``Capillary condensation''
can be understood as follows.
Consider air and liquid in a box,
where the liquid pressure $p_l$,
the total air pressure is $p_a$,
the liquid vapour pressure $p_v$,
and dry air pressure $p_{da}=p_a-p_v$.
From the Gibbs-Duhem equilibrium criterion and 
ideal gas approximation for the air, 
the Kelvin's equation \cite{Israelachvili}
is obtained:
$p_v=p_{v0}\exp\left[-(\Delta P v_l)/(R_gT)\right]$,
where $v_l$ is the partial molar volume of liquid,
$p_{v0}$ is the saturated vapour pressure,
$T$ is the absolute temperature, and $R_g$ is the gas constant.
Thus, when $\Delta P>0$, $p_v$ becomes smaller than $p_{v0}$.
Namely, in narrow spaces like pores in granular media,
the vapour can condense for lower vapour pressures.
}
Due to this effect,
the vapour can condense for lower pressures
in a narrow space, like the point where particles are in contact,
resulting in tiny liquid bridges at the contact points
between touching grains.
The critical angle $\theta_c$ increases due to 
the cohesion from these tiny bridges,
which becomes larger and increases with time
because more and more vapour condenses.
\footnote{ 
It should be noted, however, that
there are alternative explanations about the origin of 
cohesion in this experiment. 
Restagno {\it et al.} \cite{Restagno02}
conducted similar experiments for glass beads with water
and glass beads with ethanol, and found that
the difference in the increase of the critical angle 
is larger than the value expected from 
the difference of the surface tension 
between water and ethanol.
This might be partly due to the chemical reaction
between silica and water 
\cite{Restagno02,Olivi-Tran02}.
}

Hornbaker {\it et al.} \cite{Hornbaker97} 
investigated
the angle of repose $\theta_r$ of 
spherical polystyrene beads 
mixed with either corn oil or vacuum-pump oil
by the draining-crater method.
They measured the liquid content dependence of 
the angle of repose $\theta_r$.
They found that $\theta_r$ increases
linearly with the average oil-layer thickness $\delta_{\rm liq}$
(the volume of oil added divided by the total surface area of particles),
indicating that the cohesion increases as the amount of liquid increases.
However, as they increased the liquid content,
particles started to form correlated particle clusters (clumps),
and finally $\theta_r$ could not be determined in a well-defined manner.
The distribution width of the measured angle of repose 
became suddenly wide for a certain amount of liquid, 
suggesting a transition to 
a situation where long-range correlations dominate.

\paragraph{Cohesion and angle of repose}
\label{AngleofReposeThe}
\begin{figure}
\begin{center}
\includegraphics[width=0.5\textwidth]{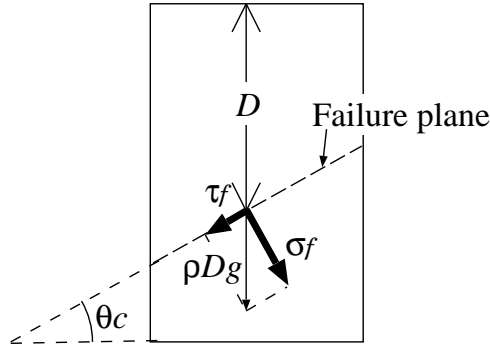}
\end{center}
\caption{
Schematic diagram of the force 
balance at the failure plane,
as determined by
the Mohr-Coulomb failure criterion.
}
\label{Fig:Coulomb}
\end{figure}
The observed increase of $\theta_r$ and $\theta_c$
with increasing liquid content
should be due to the 
cohesive force between grains.
The relation between the surface angle and cohesion, 
however, has not yet
been clearly understood.
This is an area that would greatly benefit from 
more systematic studies.

The most well-known analysis of the critical angle $\theta_c$
in granular engineering is 
to use the failure condition
for a continuum description \cite{Nedderman,Halsey98}.
In the case without cohesion, a phenomenological failure criterion
for granular media is given 
in terms of the shear stress $\tau$ and the normal compressible 
stress $\sigma$ as
\begin{equation}
\tau>\mu \sigma,
\label{Eq:CoulombCriterionNo}
\end{equation}
where $\mu$ is the internal friction coefficient,
considered as a material parameter.
If there is any plane for which the ratio $\tau/\sigma$
exceeds $\mu$, the material fails at that plane;
$\tau_f/\sigma_f=\mu$,
where the subscript $f$ for the stresses 
$\tau_f$ and $\sigma_f$ indicates the value at the failure plane.
The critical angle corresponds to the angle
of this failure plane, where the failure starts.
At the same time, because the stress comes from 
the weight above the plane, as
schematically shown in Fig.~\ref{Fig:Coulomb},
we have 
\begin{equation}
\tau_f=\rho g D \sin\theta_c \qquad \mbox{and} \qquad
\sigma_f=\rho g D \cos\theta_c,
\label{eq:taufsigmaf}
\end{equation}
with mass density $\rho$
and the depth $D$ of the failure plane from the 
surface before failure.
Namely, we have 
\begin{equation}
\frac{\tau_f}{\sigma_f}=\tan\theta_c=\mu,
\label{Eq:NonCohesiveTanc}
\end{equation}  
which gives $\theta_c$ for 
dry granular media.

In the case of wet granular media, 
it has been considered that the cohesion  
result in a normal cohesive 
pressure $\sigma_c$ \cite{Nedderman,Halsey98}.
This additional normal stress $\sigma_c$ 
allows to support a finite shear stress, even
in the limit of zero applied normal stress $\sigma$.
Then, the criterion in Eq.~(\ref{Eq:CoulombCriterionNo}), 
for non-cohesive material, is modified into \cite{Nedderman,Halsey98}
\begin{equation}
\tau > \mu (\sigma+\sigma_c).
\label{Eq:CoulombCriterion}
\end{equation}
The criterion in Eq.~(\ref{Eq:CoulombCriterion})
is called the Mohr-Coulomb criterion,
and Eq.~(\ref{Eq:CoulombCriterionNo}) is 
the special case of Eq.~(\ref{Eq:CoulombCriterion}) with $\sigma_c=0$.
With non-zero $\sigma_c$, 
the stresses at the failure plane satisfies 
$\tau_{f} = \mu (\sigma_f+\sigma_c)$,
while we still have Eq.~(\ref{eq:taufsigmaf}) for the force balance.
Then we have
\begin{equation}
\mu=\tan\theta_c\left(1+\frac{\sigma_c}{\rho g D
\cos\theta_c}\right)^{-1},
\label{Eq:CohesiveAngle}
\end{equation}
which shows that 
$\theta_c$ decreases with increasing $D$.
Namely, the criterion is 
the strictest at the bottom of the 
sandpile, which has the largest $D$.
Therefore, the failure occurs at the bottom
of the sandpile, and Eq.~(\ref{Eq:CohesiveAngle}) with
$D$ being the total depth of the sandpile
gives the critical angle $\theta_c$ of that sandpile.
Note that the relation for the non-cohesive case,
Eq.~(\ref{Eq:NonCohesiveTanc}), 
is recovered in the limit of $D\to \infty$;
the frictional force that holds the pile increases with
the size of the pile because the normal pressure increases,
but the cohesive force remains constant, 
thus the cohesive force becomes irrelevant
in a large enough granular pile \cite{Nedderman,Halsey98}.
We have increasing critical angle 
due to cohesion only for a finite sandpile
where $\sigma_c/\rho g D$ is finite,
and within this range $\theta_c$ 
increases with the cohesive stress $\sigma_c$
and depends on the pile size.
Note that, in this Mohr-Coulomb criterion,
the yield shear stress is the
consequence of the increased normal pressure,
and the direct contribution of particle-particle cohesion to
the shear stress is not taken into account.

In the case of the pendular state,
the cohesive stress $\sigma_c$
arises from the liquid bridges between particles.
Rumpf \cite{Rumpf62} proposed a simple model 
to estimate the cohesive stress 
from the force per bridge
in isotropic granular media of identical spheres with diameter $d$,
ignoring the effect of distributions in liquid bridge sizes 
and in number of bridges per particles \cite{Fournier05}.
He estimated the cohesive force $\sigma_c$ per unit area as 
\begin{equation}
\sigma_c\approx \nu \frac{k}{\pi d^2}F,
\label{Eq:RumpfModel0}
\end{equation}
where $\nu$ is the packing fraction of the grains,
$k$ is the average number of liquid bridges per particle,
and $F$ is the average force per liquid bridge.
\footnote{
In papers on granular engineering
including \cite{Rumpf62}, the porosity $n$ 
defined as the ratio of the void volume $V_v$ 
to the total volume $V_t$ is often
used instead of the packing fraction $\nu$
as a parameter to characterise the packing.
These parameters are related by
$n=1-\nu$.
}

According to the Rumpf model in Eq.~(\ref{Eq:RumpfModel0}),
$\sigma_c$ is proportional to
the cohesion per liquid bridge $F$.
Then, the fact that $\theta_r$ 
for wet spherical glass beads increases with liquid content
\cite{Hornbaker97} may suggest increasing $F$ by
increasing the liquid bridge volume,
if the number of bridge per particle $k$ stays constant.
However, this cannot be understood by 
simply considering 
the pendular state for completely spherical beads.
For the case when two spheres are in contact ($h=0$) and the wetting 
angle $\theta=0$, we get from Eq.~(\ref{eq:bridge0}) that
\begin{equation}
F_{\rm bridge}=\frac{2\pi r\gamma}{1+\tan(\phi/2)},
\label{eq:liquidbridge}
\end{equation}
where $r$ is the radius of the spheres
and $\phi$ is given in Fig.~\ref{Fig:LiquidBridge}.
Equation (\ref{eq:liquidbridge})
gives $\partial F_{\rm bridge}/\partial \phi<0$.
Namely, the attracting force $F_{\rm bridge}$
becomes smaller as the amount of liquid in the bridge increases 
while keeping the particle separation equal to zero.

This discrepancy can be resolved by considering 
the surface roughness of the grains
\cite{Albert97,Halsey98}.
It has been known that,
in the case of the cone-plane contact, 
the attracting force increases when the amount of liquid increases
\cite{Cahn70}. 
If the liquid bridge volume is small enough
to neglect the curvature of the 
macroscopic spherical shape of the beads,
the surface roughness becomes more relevant, 
and the geometry at the contact point
can be considered as the cone-plane type.
If the liquid bridge becomes wide enough
compared to the macroscopic curvature of the
surface, then the estimation for 
spheres in Eq.~(\ref{eq:liquidbridge}) is 
expected to work.

\begin{figure}
\begin{center}
\includegraphics[width=0.5\textwidth]{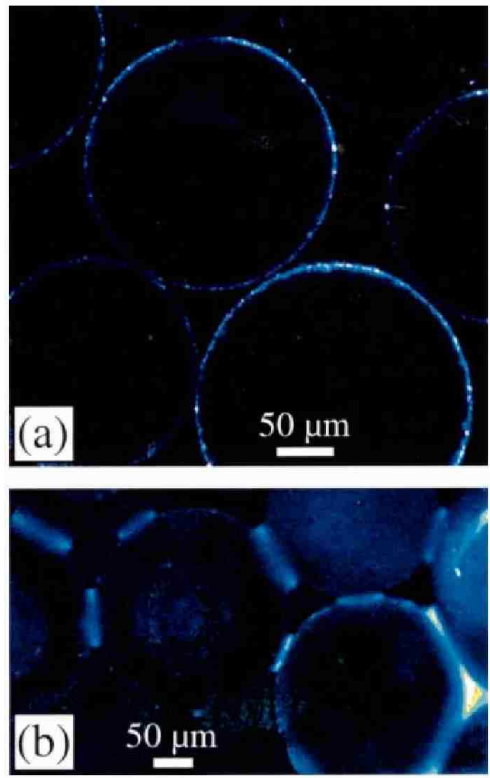}
\end{center}
\caption{
Distribution of small amount of liquid in granular media.
The blue region shows the location of the liquid.
The average grain diameter is 240 $\mu$m.
(a) 
When the amount is liquid is small, the liquid
is uniformly distributed at the grain surface.
(b) 
When the amount of liquid is large enough,
most of the liquid is caught at the contact points
between particles and liquid bridges are formed.
From 
\cite{Mason99}.
}
\label{MasonFig}
\end{figure}

Mason {\it et al.} \cite{Mason99} experimentally observed
how the liquid distributes in spherical beads 
with microscopic surface roughness.
They found that, when the amount of liquid is 
smaller than a critical value,
most of the liquid is trapped on the particle surface due 
to the surface roughness 
(Fig.\ref{MasonFig}(a)) 
and the critical angle did not depend on the surface tension of the
liquid.
When the liquid content exceeds a critical value,
liquid bridges are formed at the contact points 
(Fig.\ref{MasonFig}(b)), and the critical angle increases 
with liquid volume and depends on the liquid surface tension.
This agrees with the behaviour expected from 
the regime where the surface roughness 
determines the cohesion from a liquid bridge.

However, one should note that,
not only the cohesion force $F$ but also
the number of liquid bridges $k$ per particles 
depends on the liquid content in general, 
which should also affect the cohesive stress $\sigma_c$.
Recent experiments \cite{Kohonen04,Fournier05} show that 
$k$ increase very rapidly upon liquid content 
for small liquid content, and then saturate. 
The effect of $k$ on the critical angle needs to be investigated
carefully especially for small liquid content.

Albert {\it et al.} 
\cite{Albert97} proposed a criterion to understand the critical
angle different from the Mohr-Coulomb criterion. 
They considered the criterion whether
a grain at the surface escapes from
the bumpy geometry that other grains form.
In their theory, the surface grain 
at an unstable position rolls down
and the final angle is determined by 
the condition that all grains 
at the surface sit at stable positions. 
The cohesion from liquid 
affects the force balance and 
increases the critical angle $\theta_c$.

In either mechanism,
the Mohr-Coulomb criterion or the surface 
failure criterion, 
the angle of repose of cohesive granular media is larger than
that of non-cohesive granular media. 
However, the location of the failure plane is different.
The failure occurs at the bottom 
in the case of the Mohr-Coulomb criterion
and $\theta_c$ depends on the system size,
while the surface fails for the ``grain escape'' criterion
and $\theta_c$ is independent of the system size.
In addition, the Mohr-Coulomb criterion is 
a continuous description, 
while the grain escape criterion explicitly considers 
the discreteness of the system. 

\begin{figure}
\begin{center}
\includegraphics[width=0.7\textwidth]
{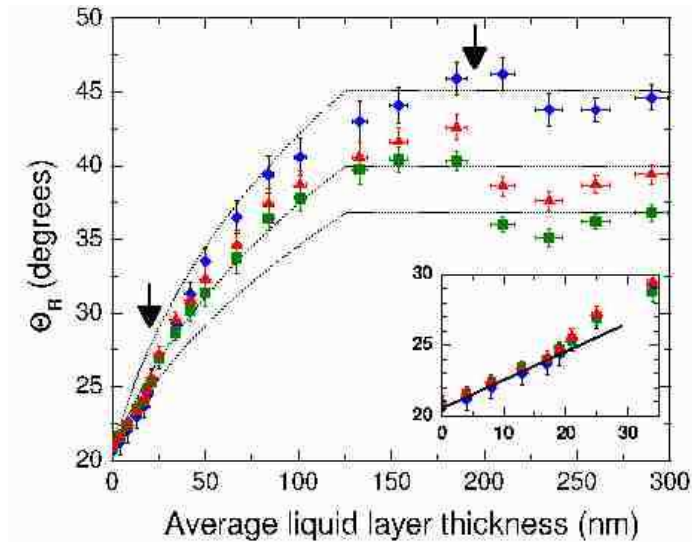}
\end{center}
\caption{
Angle of repose $\theta_r$
versus the average oil-layer thickness $\delta_{\rm liq}$
for spherical glass beads with average diameter 0.9mm
with three different container diameters $d_c$
(diamonds: 10.3cm, 
triangles: 15.6cm, 
squares: 20.4cm).
The inset shows an enlargement of small
$\delta_{\rm liq}$ regime.
The two vertical arrows indicate the 
transitions between the granular, 
correlated, and plastic regime.
From \cite{Tegzes99}. 
}
\label{Fig:Tegzes99Fig2}
\end{figure}

Both mechanisms may be responsible in determining
the critical angle. However, it is likely that
the discrete picture is dominant for
very small amounts of liquid content,
with small cohesive bulk stress,
while the continuous picture is appropriate when
the liquid content is enough to cause the cohesive bulk stress
comparable to gravity.
In order to see the relevant mechanism,
Tegzes {\it et al.} \cite{Tegzes99} examined 
the system size dependence of the angle of repose $\theta_r$
for a larger range of liquid content (Fig.~\ref{Fig:Tegzes99Fig2}). 
They adopted the draining crater method,
and the system size is controlled
by changing cylinder diameters.
They found that $\theta_r$ is independent of system size
for small enough liquid content 
(called the {\it granular regime}, where 
surface flow occurs at the top few layers),
while it decreases with system size for
larger liquid content
(called the {\it correlated regime}, where 
grains form clumps and the surface flow is correlated).
At the largest liquid content, the angle 
first decreases and then increases slightly with 
oil-layer thickness, but still depends on the system size
(called the {\it plastic regime}, where
the medium retains a smooth crater surface 
and the motion is coherent).
Qualitative agreements have been obtained 
with the surface failure criterion 
for the granular regime
and with the Mohr-Coulomb criterion for the correlated and plastic
regime.

These results are not yet conclusive,
and many researches on the surface angle are still going on.
Ertas {\it et al.} \cite{Ertas02} considered the critical 
angle in a pile made of a
mixture of spheres and dumbbell-shape grains,
extending the surface failure criterion by \cite{Albert97}.
They found that the surface angle drastically increases
as the fraction of dumbbell-shape grains increases,
because the dumbbell-shape grains hardly roll down the slope.
They pointed out that, in wet granular media
with small liquid content,
liquid bridges connect a few grains and make clumps;
these clumps may contribute to the increasing surface angle
through the same mechanism that the fraction of 
dumbbell-shaped grains increased the critical angle.
On the other hand, Nowak {\it et al.} \cite{Nowak05} 
applied the geometrical consideration of force balance
by \cite{Albert97}
not only at the surface but also inside the material.
They claimed that the most unstable plane from 
the geometrical consideration is inside the material.
Their argument does not take into account the effect of friction,
but it gives system-size dependent critical angle
that agree with the critical angle of dry grains in 
the limit of infinite system size;
this feature is similar to 
that in the Mohr-Coulomb criterion.

In summary, the effect of cohesion on surface angle
in granular pile has not yet been clearly understood,
and further systematic studies are needed.

\subsubsection{Tensile, compression, and shear tests
for intermediate and large liquid content}
\label{tests}
For large enough cohesion,
the angle of repose may exceed $90^\circ$
for laboratory-scale granular piles. In such a regime,
the angle of repose is no longer a 
good parameter to characterise the cohesion,
and we need another method to characterise the cohesion 
in wet granular media.

If the failure condition is characterised by
the Mohr-Coulomb criterion (\ref{Eq:CoulombCriterion}),
the measurement of the stresses at the 
failure gives the cohesive stress $\sigma_c$,
which characterises the cohesion in wet granular media.
The analysis of granular media based on the
Mohr-Coulomb criterion has been well-established 
in the engineering field, 
and a considerable amount of data has been accumulated 
for wet granular media.
In this subsection, we briefly describe several 
experimental methods to 
determine the internal friction coefficient
$\mu$ and the cohesive stress $\sigma_c$, 
and summarise what has been known for
wet granular media from those measurements.

\paragraph{Test methods and Mohr circle}
\begin{figure}
\begin{center}
\includegraphics[width=0.8\textwidth]{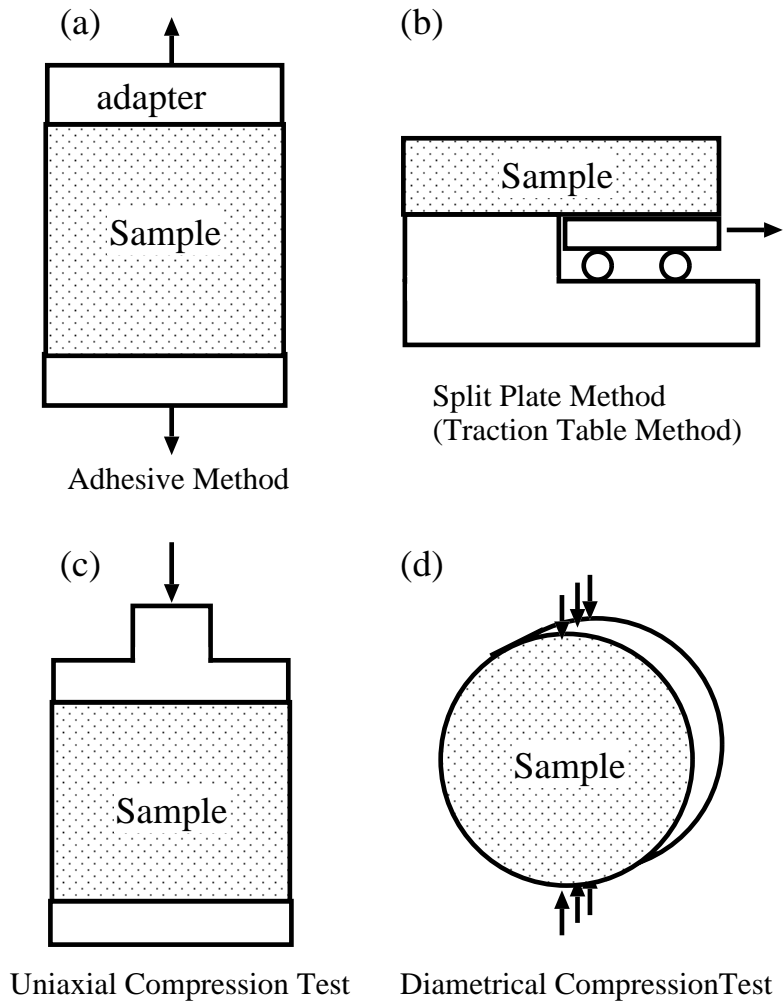}
\end{center}
\caption{
Several measurement techniques of strength of granular media.
The adhesive method (a),
the split-plate method (traction table method) (b),
the uniaxial compression test (c),
and the diametrical compression test (d).
}
\label{Fig:Schubert75-2Fig4}
\end{figure}
Several experimental methods have been used to measure 
the tensile strength of wet granular media
(e.g., \cite{Bika01,Schubert75-2}).
The simplest method is just to pull the sample apart until the
material fails, as shown 
in Figs.~\ref{Fig:Schubert75-2Fig4}(a) and (b).
However, it is difficult to hold the sample
because wet granular media are fragile and not always uniform.
In the adhesive method (Fig.~\ref{Fig:Schubert75-2Fig4}(a)), 
the material needs to be strong enough so
that the bottom and top faces of the sample
can be bonded to the adaptors.
In the case of the split-plate method
(Fig.~\ref{Fig:Schubert75-2Fig4}(b)), 
the sample is placed on a plate,
and the plate is split until the material fails.
In either method, 
both the stress and the strain can be measured until reaching failure.

The uniaxial and diametrical compression tests have also been 
conducted, where the sample is compressed
in one direction until it fails.
In uniaxial compression tests
(Fig.~\ref{Fig:Schubert75-2Fig4}(c)),
a sample is pressed by a ram along one direction.
In the case of the diametrical compression test,
a linear load is applied to a sample of cylindrical shape
across one diameter of the disk 
(Fig.~\ref{Fig:Schubert75-2Fig4}(d)),
and the tensile stress is calculated 
based on the Hertz theory 
for isotropic elastic bodies \cite{Procopio03,Hertz},
though the validity of the
elastic theory for granular media 
with the assumption of the isotropy
is rather suspicious.

\begin{figure}
\begin{center}
\includegraphics[width=.9\textwidth]{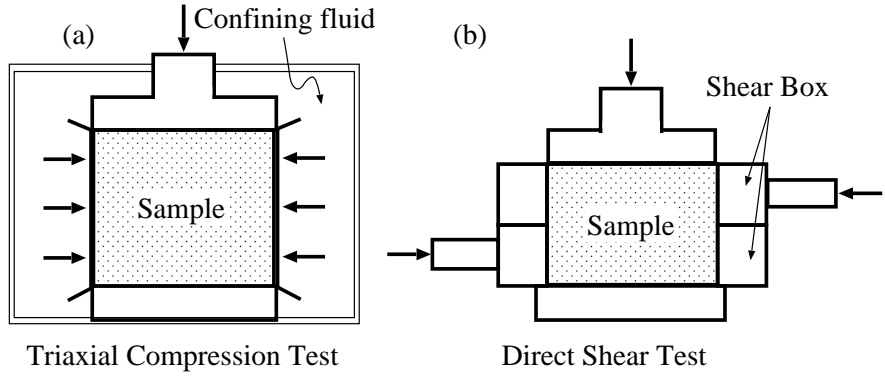}
\end{center}
\caption{Schematic diagrams of 
a triaxial test system (a)
and a direct shear test system (b). 
}
\label{Fig:SoilTests}
\end{figure}

In addition to these uniaxial tests, where 
only one component of the stress applied to the sample is controlled,
there are methods where two components 
of the stress are separately controlled. 
The triaxial compression test
has been widely performed for soils \cite{SoilBook,UnsaturatedBook,
SoilJapaneseBook}.
Figure \ref{Fig:SoilTests}(a) shows a 
schematic description of
the triaxial testing system.
The cylindrical-shaped sample is placed in a confining cell
filled with fluid, and the sample is separated 
from the confining fluid by a flexible membrane. 
The sample is compressed from above by a ram, 
and the pressure at side membrane can be 
separately controlled
by changing the pressure of the confining fluid.
In the case of the direct shear test
in Fig.~\ref{Fig:SoilTests}(b),
a sample in a shear box is sheared as the shear box is slid.
A series of tests can be 
conducted by varying the vertical compressive stress 
from the top.
Note that there are several modifications of these tests:
In the case of unsaturated soils, 
the modified tests are often performed,
where the suction is kept constant during the test
by using the HAE material (see subsection \ref{suctionmeasure})
and allows the liquid to drain from the sample \cite{UnsaturatedBook}.

It should be noted that the sample needs to be
prepared carefully in either method
in order to get reproducible results. 
Especially, how the sample is compacted 
before the measurement is an important factor to 
determine its mechanical response.

\begin{figure}
\begin{center}
\includegraphics[width=0.9\textwidth]{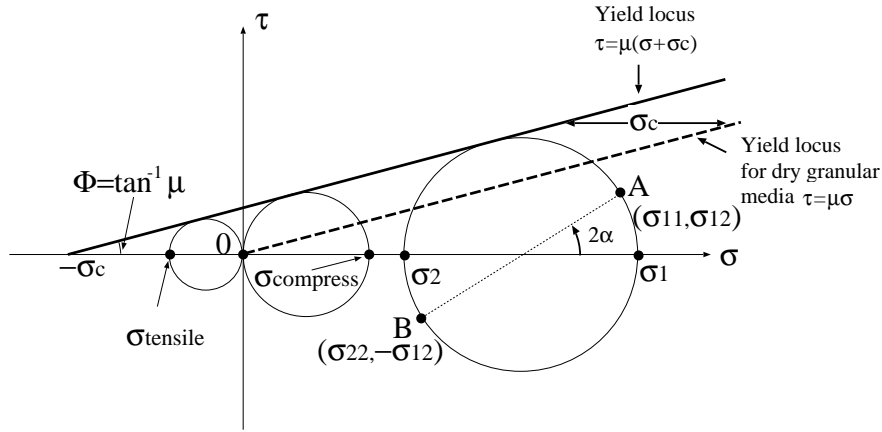}
\end{center}
\caption{
A schematic diagram of 
Mohr circles in the plane 
shear stress $\tau$ versus normal compressive stress $\sigma$. 
When a two-dimensional stress tensor 
(whose $(i,j)$ component is given by $\sigma_{ij}$)
is considered, the Mohr circle includes the diametrically-opposed points 
$A (\sigma_{11}, \sigma_{12})$
and $B (\sigma_{22}, -\sigma_{12})$,
as shown in the figure.
The yield locus from the Mohr-Coulomb criterion 
for cohesive granular media is shown by the solid straight line,
and the yield locus of the non-cohesive granular media 
is shown by the dashed line.
}
\label{Fig:MohrCircle}
\end{figure}
The results of the tensile, compression, and shear tests are often analysed 
by using the Mohr circle, which
visualises the rotational transformation of 
the two-dimensional stress tensor as follows
\cite{SoilBook,UnsaturatedBook,SoilJapaneseBook}. 
Let us consider a two-dimensional stress tensor
whose principal values are $\sigma_1$ and $\sigma_2$.
If the coordinate axes are rotated by an angle $\alpha$ from the
principal axes, the stress tensor $\sigma_{ij}$ in 
the coordinate system is given by
\begin{eqnarray}
\left(
\begin{array}{cc}
\sigma_{11} & \sigma_{12}\\
\sigma_{21}&\sigma_{21}
\end{array}
\right)
&=&
\left(
\begin{array}{cc}
\cos\alpha & \sin\alpha \\
-\sin \alpha &\cos \alpha 
\end{array}
\right)
\left(
\begin{array}{cc}
\sigma_{1} & 0\\
0 &\sigma_{2}
\end{array}
\right)
\left(
\begin{array}{cc}
\cos\alpha & -\sin\alpha \\
\sin \alpha &\cos \alpha 
\end{array}
\right)\nonumber \\
&=&
\left(
\begin{array}{cc}
\frac{(\sigma_1+\sigma_2)}{2}+\frac{(\sigma_1-\sigma_2)}{2}\cos
2\alpha&
\frac{(\sigma_1-\sigma_2)}{2}\sin2\alpha\\
\frac{(\sigma_1-\sigma_2)}{2}\sin2\alpha&
\frac{(\sigma_1+\sigma_2)}{2}-\frac{(\sigma_1-\sigma_2)}{2}
\cos 2\alpha
\end{array}
\right)
\end{eqnarray}
Namely, when we consider two points $A (\sigma_{11}, \sigma_{12})$
and $B (\sigma_{22}, -\sigma_{21})$ 
in the $\tau-\sigma$ plane,
they locate at the diametrically-opposed points 
on a circle
\begin{equation}
\left[\sigma-\frac{1}{2}(\sigma_1+\sigma_2)\right]^2+\tau^2
=\left[\frac{1}{2}(\sigma_1-\sigma_2)\right]^2,
\end{equation}
as shown in Fig.~\ref{Fig:MohrCircle},
where the angle between the $\sigma$ axis and
the line $A-B$ is $2\alpha$.

The Mohr circle is useful when considering the
Mohr-Coulomb criterion Eq.~(\ref{Eq:CoulombCriterion}).
The stress at failure, from the Mohr-Coulomb
criterion, gives a straight line 
\begin{equation}
\tau=\mu(\sigma+\sigma_c)
\label{tangentline} 
\end{equation}
in the $\tau$-$\sigma$ plane
with $\sigma$-intercept at $-\sigma_c$
and slope $\mu$.
Instead of $\mu$, 
$\Phi=\tan^{-1}\mu$ (shown in Fig.~\ref{Fig:MohrCircle})
is also used as a parameter for the
Mohr-Coulomb criterion; 
$\Phi$ is called the internal friction angle.
If the granular aggregate obeys the Mohr-Coulomb criterion,
the Mohr circle at failure should be
tangent to this straight line (\ref{tangentline}),
because the granular aggregate fails as soon as 
a stress state that satisfies the criterion appears.
Therefore, by drawing an envelope curve 
of Mohr circles at failure
with various stress conditions,
from the data one can construct the Mohr-Coulomb 
criterion to determine 
the parameters $\mu$ and $\sigma_c$.

This procedure can be done for the triaxial compression test 
and the shear test, where 
two components of the stress tensor are 
varied separately.
The granular material does not always 
follow the ideal Mohr-Coulomb criterion
and the envelope might be curved, but
the response is linear
for small enough values of the stress,
and this gives $\mu$ and $\sigma_c$.

In the case of uniaxial tensile (compression) tests, 
one of the principal values is always zero.
Thus, the resulting Mohr circle crosses the origin 
of the $\sigma$-$\tau$ plane, 
where the tensile (compressive) stress
at failure is given by $\sigma_{\rm tensile}$
($\sigma_{\rm compress}$) as shown in Fig.~\ref{Fig:MohrCircle}.
For a given internal friction coefficient 
$\mu$, $\sigma_{\rm tensile}$ ($\sigma_{\rm compress}$) 
is proportional to $\sigma_c$,
and the behaviour of
$\sigma_{\rm tensile}$ ($\sigma_{\rm compress}$) 
gives information about the cohesion in the wet granular medium.
However, we cannot determine the cohesive stress $\sigma_c$ from 
a uniaxial test unless we know $\mu$ from another experiment.

Below let us briefly review some experimental results
about the stresses at the failure state
for wet granular media with various liquid content.

\paragraph{Tests in the pendular state}
Pierrat {\it et al.} \cite{Pierrat98} investigated
the Mohr-Coulomb criterion by using direct shear tests
for various granular assemblies
including monodisperse glass beads 
and polydisperse crushed limestone
with relatively small liquid content. 
The amount of liquid was larger than that in
the experiments on the angle of repose presented 
in subsection \ref{AngleofRepose}, 
but the system was supposed to be still in the pendular state.

\begin{figure}
\begin{center}
\includegraphics[angle=-90,width=0.6\textwidth]{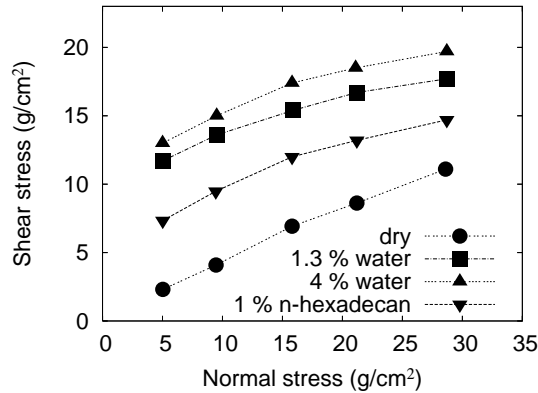}
\end{center}
\caption{
Yield locus obtained by
direct shear tests for glass beads (93 $\mu$m):
Dry (circles), 
with water (1.3\% of liquid content is shown by squares,
and 4\% of liquid content is shown by triangles),
and with n-hexadecan 
(liquid content 1\%, shown by upside-down triangles).
Adapted from 
\cite{Pierrat98}. 
}
\label{Fig:Pierrat98Fig4}
\end{figure}

The yield loci in shear test for glass beads of
diameter 93 $\mu$m with various liquid content
are shown in Fig.~\ref{Fig:Pierrat98Fig4},
which shows a shift of the yield locus to the left 
from the dry case,
but the slope is unchanged.
It has also been found that the shift increases 
with the liquid content.
For all the materials they investigated,
they found that the yield loci 
can be collapsed by shifting the curve.
Namely, in this experiment, the main effect of the
liquid appears in the cohesive stress $\sigma_c$,
but the internal friction coefficient $\mu$ 
was not significantly modified,
which indicates that the effect of lubrication
was small.

They also investigated the relation 
between the shear test and the uniaxial tensile test.
They had conducted tensile tests separately
\cite{Pierrat97}, 
and from their empirical relation
between $\sigma_{\rm tensile}$  and parameters that characterise the
sample, such as packing fraction and liquid content,
they estimated the tensile strength $\sigma_{\rm tensile}$ 
for the sample in the shear tests. 
If the internal friction coefficient $\mu=\tan\Phi$ for 
dry samples does not change for wet granular materials,
we can estimate $\sigma_c$ from 
$\sigma_{\rm tensile}$, using the relation
$\sigma_c=(1+\sin\Phi)|\sigma_{\rm tensile}|/(2\sin\Phi)$
(see Fig.~\ref{Fig:MohrCircle})
with $\Phi$ being the internal friction angle in the dry case.
This estimation was compared 
with observed yield loci in shear tests,
and reasonable agreement was found.

\paragraph{Tests with intermediate liquid content}
Next, we review
the tensile tests and uniaxial compression tests
of wet granular media with intermediate liquid content,
typically in the funicular state.
The mechanical response of the funicular state 
is not well understood,
because in this state 
even the qualitative dependence 
of the tensile strength 
(tensile stress at the peak of stress-strain relation)
on liquid content
varies from one material to another,
and the reason for this is not very clear. 
Here we just summarise a few experiments.

\label{granules}
\begin{figure}
\begin{center}
\includegraphics[width=0.6\textwidth]{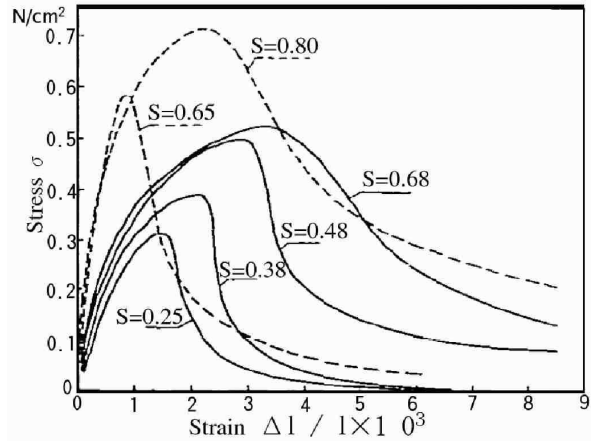}
\end{center}
\caption{
Stress-strain relation of wet limestone (mean diameter 65 $\mu$m)
obtained by the adhesive method \cite{Schubert75}.
The packing fraction $\nu$ of the sample was $0.66$.
The dashed lines show the result
for a sample prepared by a drying process (drainage), 
while the solid lines show the result
for a sample prepared by a wetting process (imbibition).
$S$ refers to the percentage liquid content.
From 
\cite{Schubert75}. 
}
\label{Fig:Schubert75Fig9}
\end{figure}
\begin{figure}
\begin{center}
\includegraphics[width=0.6\textwidth]{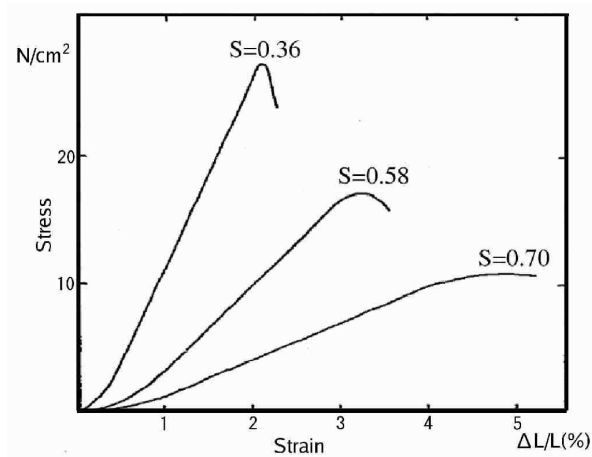}
\end{center}
\caption{Stress-strain relation of
dicalcium phosphate (diameter 21 $\mu$m)
with an aqueous solution of a polymer.
$S$ refers to the percentage liquid content.
The packing fraction of the sample was $0.5$.
From 
\cite{Holm85}. 
}
\label{Fig:Holm85Fig8.ps}
\end{figure}

In the field of agglomeration processing, experiments 
have been conducted from relatively dry to
almost saturated conditions.
However, in the funicular state,
the liquid content dependence of
the strength may either increase
or decrease upon increasing the liquid content.
For example,
increasing tensile strength
with increasing liquid content has
been found by 
Schubert \cite{Schubert75} 
for limestone with mean diameter 65$\mu$m
using the adhesive method.
Figure \ref{Fig:Schubert75Fig9} shows
the stress-strain relation for various liquid content $S$,
and the peak stress gives tensile strength.
Stress-strain curves, showing
decreasing strength
as the liquid content $S$ increases from 36\% to 70\% 
are shown in Fig.~\ref{Fig:Holm85Fig8.ps},
from 
\cite{Holm85}, where 
compressive tests were performed using
dicalcium phosphate with diameter 21 $\mu$m.
Kristensen {\it et al.} \cite{Kristensen85} found increasing strength
for increasing liquid content using glass beads of particle size 68 $\mu$m,
while  decreasing strength
for dicalcium phosphate of particle size 14 $\mu$m in 
uniaxial compression tests.
There have been many other experiments probing the
mechanical strength of wet granular media 
(see \cite{GranulationReview} for a review). 
The effect of particle size on the competition 
between cohesion and lubrication was considered \cite{GranulationReview} 
to be one of the causes of these various behaviours,
but the liquid content dependence in the 
intermediate liquid-content-regime 
(approximately given by $20 \% \lesssim S \lesssim 90 \%$)
is not yet well-understood.

It has been found that the critical strain $\epsilon_c$
(strain at the peak in the stress-strain curve)
always increases with increasing liquid content,
as shown in Figs.~\ref{Fig:Schubert75Fig9} and \ref{Fig:Holm85Fig8.ps}.
This indicates that the wet granular material
is brittle when the amount of liquid is small,
and tends to show visco-plastic behaviour as the liquid content is increased.

It should be also noted that,
even for similar amounts of liquid content,
the strength of a sample may depend on whether it is 
prepared by either draining or adding liquid;
Schubert \cite{Schubert75} found different 
stress-strain curves for each processes 
as shown in Fig.~\ref{Fig:Schubert75Fig9}.
In many studies, the sample is often driven by an external force
(e.g., rotating drum, mixing, etc.) before the measurement
in order to better distribute
the liquid, which would make hysteresis less obvious.

\paragraph{Tests in soil mechanics:
relatively large amounts of liquid}
\begin{figure}
\begin{center}
\includegraphics[angle=-90,width=0.6\textwidth]{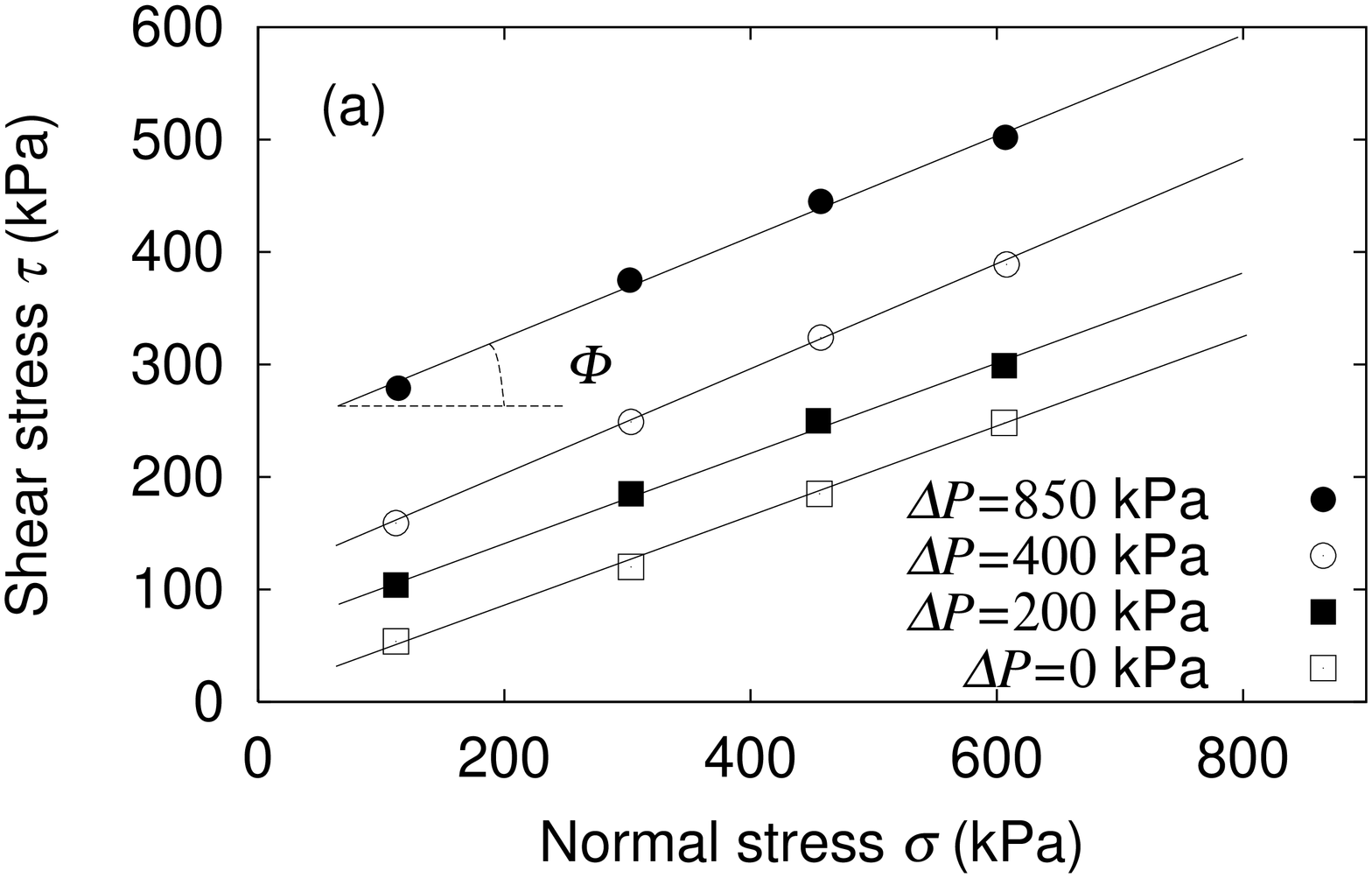}
\includegraphics[angle=-90,width=0.6\textwidth]{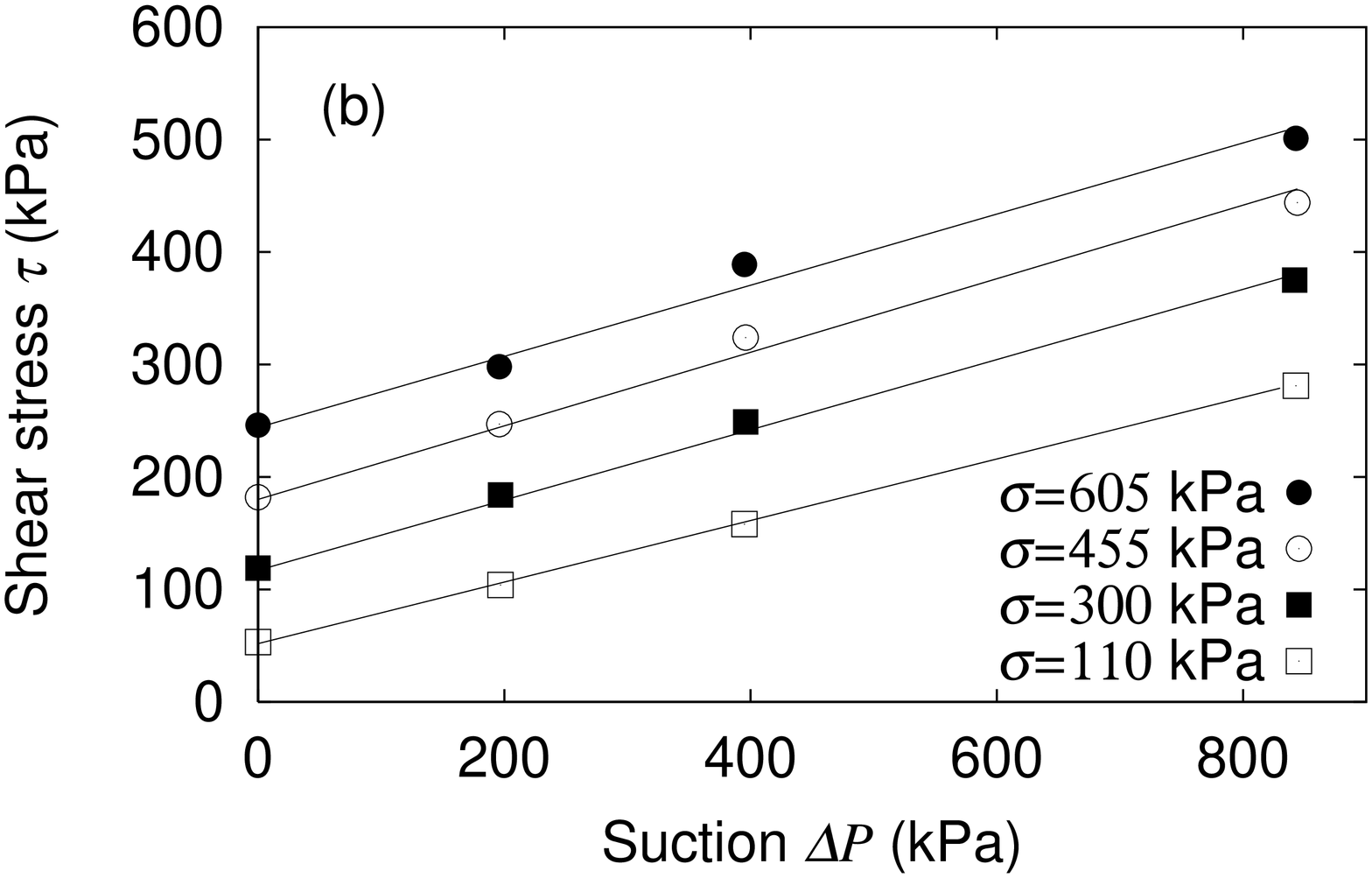}
\end{center}
\caption{Yield loci for wet soil (Madrid Gray Clay).
(a) Net normal stress $\sigma$ versus shear stress $\tau$ for
various suction $\Delta P=P_a-P_l$.
The slope is almost independent of suction in this regime.
(b) Shear stress versus suction for various
net normal stresses $\sigma$.
The slope is almost independent of the net normal stress in this regime.
Adapted from 
\cite{UnsaturatedBook} 
(original data from 
\cite{Escario80}).
}
\label{Fig:UnsaturatedBookFig6-6}
\end{figure}

\begin{figure}
\begin{center}
\includegraphics[angle=-90,width=0.6\textwidth]%
{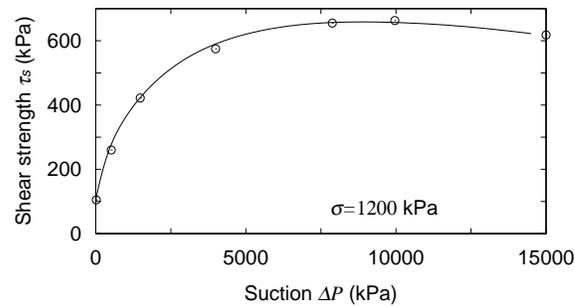}
\end{center}
\caption{Shear strength $\tau_s$ versus suction $\Delta P$
obtained by shear tests 
for red clay. 
This figure shows a
larger range of suction $\Delta P$
(compared to Fig.~\ref{Fig:UnsaturatedBookFig6-6}(b)),
as well as a non-linear and non-monotonic behaviour.
Adapted from 
\cite{Escario89}.
}
\label{Fig:UnsaturatedBookFig6-12}
\end{figure}

As pointed out in subsection \ref{phases},
the suction in the capillary state largely
varies upon a small change of liquid content.
In soil mechanics, the shear tests
of unsaturated soils have been conducted
by decreasing the water content from initially saturated soils,
or, in other words, 
by increasing the suction from zero.
The data is usually presented in terms of the suction $\Delta P$,
not the liquid content $S$.
This is a reasonable way of arranging data in the capillary state,
because the change of $\Delta P$ upon changing $S$ is 
very rapid for low and high liquid content 
as has been shown in subsection \ref{phases}.
Here we summarise these data to try to understand the 
mechanical properties with large amount of liquid content.
It should be noted that the grains
that form soils can be 
very small (for example, clay can be the order of 
2 $\mu$m in diameter), may have characteristic shape,
and may be affected by, e.g., electrostatic interaction. 
These properties affect its mechanical responses,
but we focus on the variation of 
the mechanical properties upon changing the suction $\Delta P$
in the following, so that
we will be able to extract the properties 
mainly determined by the capillary effect.

The failure condition given by
the Mohr-Coulomb criterion
in Eq.~(\ref{Eq:CoulombCriterion})
has been experimentally investigated for soils.
Figure \ref{Fig:UnsaturatedBookFig6-6}(a)
shows\footnote{
In soil mechanics, the total normal stress $\sigma_t$,
the air pressure $P_a$, and the water pressure $P_l$
are often taken to be the control parameters.
Here, the total normal stress $\sigma_t$ 
includes both the load on the granular sample
and the fluid pressure on it, thus
$\sigma_t=P_a$ if the sample is 
placed in air without any external load.
When an external pressure is applied on the sample,
the net normal stress $\sigma$ 
sustained by granular particles would be given by $\sigma_t-P_a$.
In the original figures of Fig.~\ref{Fig:UnsaturatedBookFig6-6}(a),
the label of the horizontal axis shows
``Net Normal Stress $\sigma-u_a$ (kPa)'',
where $\sigma$ in \cite{UnsaturatedBook} represents the total normal stress,
and $u_a$ is the air pressure ($\sigma_t$ and $P_a$ in our notation,
respectively).
}
the yield loci of wet clay for various 
values of the suction $\Delta P$,
where the slope gives the friction coefficient $\mu$
and the $y$-intercepts of lines give $\mu\sigma_c$.
The slopes scarcely vary with suction,
which indicates that $\mu$, 
or, the internal friction angle $\Phi=\tan^{-1}\mu$,
does not depend on the suction $\Delta P$.
The $y$-intercept is small for $\Delta P=0$ and 
increases with increasing the suction $\Delta P$.
Namely, $\sigma_c$ increases with increasing $\Delta P$.

Figure \ref{Fig:UnsaturatedBookFig6-6}(b) 
shows the yield shear stress versus suction 
for various normal compressive stresses,
which shows a linear dependence on $\Delta P$.
This result and the fact that the 
internal friction angle $\Phi$ 
depends little on the suction $\Delta P$
suggest that the cohesive stress $\sigma_c$
increases linearly with $\Delta P$. 
The linear increase of 
$\sigma_c$ 
upon increasing $\Delta P$ for small $\Delta P$
is natural, because the compressive 
stress $\Delta P$ at the liquid-air interface
at the surface of the granular assembly
is the source of the cohesion.
This behaviour has been found 
for some kinds of clay \cite{Escario80,Escario89,Bishop60},
undisturbed decomposed granite \cite{Ho82},
silt \cite{Krahn89},
and glacial till \cite{Gan88}.

When $\Delta P$ is increased further,
a non-linear dependence of the shear strength $\tau_s$
(the shear stress at failure for a given suction and normal load)
on $\Delta P$ is found \cite{Escario89,UnsaturatedBook}, as shown in 
Fig.~\ref{Fig:UnsaturatedBookFig6-12}. 
The increase of the shear strength $\tau_s$ becomes 
nearly zero as $\Delta P$ grows.
It is very likely that the internal friction coefficient $\mu$ 
does not vary significantly upon changing $\Delta P$ in these regimes,
and then the obtained 
shear strength $\tau_s$ is proportional to the cohesive stress $\sigma_c$. 

Summarising these results, 
$\sigma_c$ increases linearly 
with $\Delta P$ for small enough suction $\Delta P$,
but for large $\Delta P$, the increase disappears.
The suction $\Delta P$ increases as the liquid content $S$ 
decreases as shown in Fig.~\ref{Fig:SoilWater},
therefore, 
$\sigma_c$ increases upon decreasing the liquid content
from $S=100\%$ ($\Delta P=0$)
\cite{Vanapalli96}.

\subsection{Dynamical behaviours}
\label{dynamicalbehaviours}
In this subsection, we briefly describe some dynamical
behaviours of wet granular media. 
In the dynamical regime, not
only cohesion but also other effects of the liquid
 such as viscosity, lubrication, and liquid motion
play important roles.
The dynamical behaviours observed
are far more complicated
than those in the quasistatic regime.
In addition, though there are numerous
studies of the dynamics of partially wet granular media 
from the practical point of view,
systematic studies 
in simple situations are rather few;
here we briefly summarise some of the experimental studies
on the dynamical response of wet granular media.
Those who are interested in recent studies on 
dynamics should also refer to a recent review
\cite{Herminghaus05}.

\subsubsection{Dynamics in the pendular state}

\paragraph{Avalanches in rotating drums}
\label{Tegzes}
\begin{figure}
\begin{center}
\includegraphics[width=0.8\textwidth]{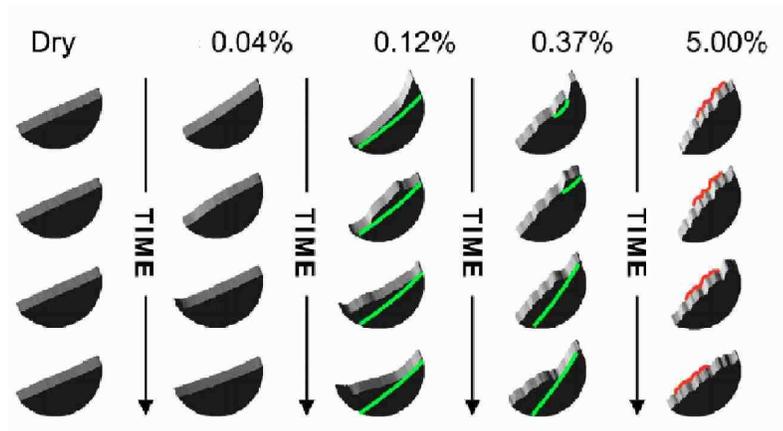}
\end{center}
\caption{
Surface shapes in a rotating drum 
during single avalanches, for various liquid contents $S$
(shown as percentages).
The coloured line for liquid content percentages $0.12\%$ 
and 0.37\% indicate approximate slip planes in the correlated regime, 
and the coloured lines for $5.00\%$ 
show the travelling quasiperiodic surface features 
in the viscoplastic regime.
From 
\cite{Tegzes03}. 
}
\label{Fig:Tegzes03Fig17a}
\end{figure}
Granular flow and avalanches in a rotating drum
have been widely investigated for dry granular media
(see, e.g., 
\cite{Rajchenbach90,Rajchenbach02,Bonamy02,RajchenbachReview,
Benza93,Ristow96,Khakhar97,Baumann95}).
At low rotation rate, the grains stay at rest
until the surface angle reaches the 
critical angle $\theta_c$,
and avalanche occurs;
the angle just after the avalanche is 
the angle of repose $\theta_r$.
This is called the intermittent regime.
When the rotation rate is high enough, 
the continuous flow occurs with keeping 
a constant angle between the surface and horizontal plane;
this is the continuous regime.
The flow regimes of fine cohesive powders in a rotating drum has
been also studied by \cite{Castellanos99}.

Tegzes {\it et al.} \cite{Tegzes02,Tegzes03} investigated
the flow behaviours of wet granular materials
in a thin rotating drum. In the experiments,
the amount of liquid is small 
(liquid content varied from 0.001\%
to 5\%),
where the pendular state is expected.
They have observed 
the transition from intermittent to continuous flow
with increasing the rotation rate
in this range of liquid content.

They observed the time evolution of the surface profile,
which enable them to calculate the mean surface angle,
statistics of the local surface angle,
evolution of pattern at the surface, etc.
In Fig.~\ref{Fig:Tegzes03Fig17a},
the typical surface profiles
just after avalanches in the intermittent flow regime are 
shown. For enhanced clarity, the gray-scale plot in the 
third dimension of the figure shows
the local surface angles.

By changing the liquid content,
the following three regimes have been distinguished. 
For low liquid content, the cohesion is small,
and small avalanches are observed (the {\it granular} regime);
the avalanches occur at the surface and
the surface profile is always smooth
(Fig.~\ref{Fig:Tegzes03Fig17a}, liquid content $0.04\%$), 
similarly to the dry granular media
(Fig.~\ref{Fig:Tegzes03Fig17a}, dry case).
When increasing the liquid content, 
the cohesion becomes stronger,
and grains move as a connected block.
An avalanche occurs
through a succession of local slip events,
and the surface structure fluctuates after
each avalanche; this is the {\it correlated} regime
(Fig.~\ref{Fig:Tegzes03Fig17a}, liquid content $0.12\%$ and $0.37\%$).
Further increasing the liquid content
results in the {\it visco-plastic} regime.
In this state, the flow becomes
coherent over the entire sample,
and fluctuations are suppressed 
(Fig.~\ref{Fig:Tegzes03Fig17a}, liquid content $5\%$).

They also found that 
the critical angle $\theta_c$ depends
not only on the liquid content but also on the rotation rate.
They varied the duration between avalanches, 
or the ``waiting time'', from about 0.1 s to 1000 s. 
It was found that $\theta_c$ increased logarithmically 
with the waiting time.
They supposed that this was caused by the slow 
flow of liquid along the particle surface.

A systematic study of liquid motion in 
wet granular media in the pendular state 
was recently conducted by \cite{Kohonen04}.
They found that, just after shaking the material to 
mix the liquid with grains,
the average liquid bridge volume was much less than that 
in equilibrium;
the volume saturated after long enough waiting times,
more than one hour.
This also suggests that the liquid-moving process
is very slow and may induce the ageing effect
observed by \cite{Tegzes02,Tegzes03}.

\paragraph{Vibrated wet granular media}
\label{vibratedwet}
Dry granular media placed on a horizontal plane
excited by vertical vibrations
is a fundamental and widely-used setup 
to study granular gases (e.g., 
\cite{GranularGas}),
segregation (e.g., 
\cite{GranularSeg}),
and pattern formation
(e.g., 
\cite{GranularPattern}).
In dry granular media,
the onset of fluidisation
is often characterised by the dimensionless
acceleration $\Gamma=A\omega^2/g$,
where $A$  is the amplitude of the vibration, $\omega$
is the angular frequency of the vibration,
and $g$ is the acceleration of gravity.
There is a threshold $\Gamma_{\rm min}$ below which 
there are no fluidisation, which was reported to be a constant
around one, (e.g., \cite{Duran97,Clement92,Knight93}),
but recently reported that
$\Gamma_{\rm min}$ can be smaller than one
and show weak $\omega$ dependence \cite{Renard01,Poschel00}. 
Above the fluidisation threshold,
dry granular media exhibits 
transitions between localised patterns of jumping grains 
(in the case of few layers of grains)
all the way to gas-like phases 
(see, e.g.,
\cite{Losert99,Melo95,DAnna03,
Umbanhowar96,Pak94,Laroche89,DAnna01}).

\begin{figure}
\begin{center}
\includegraphics[width=0.48\textwidth]{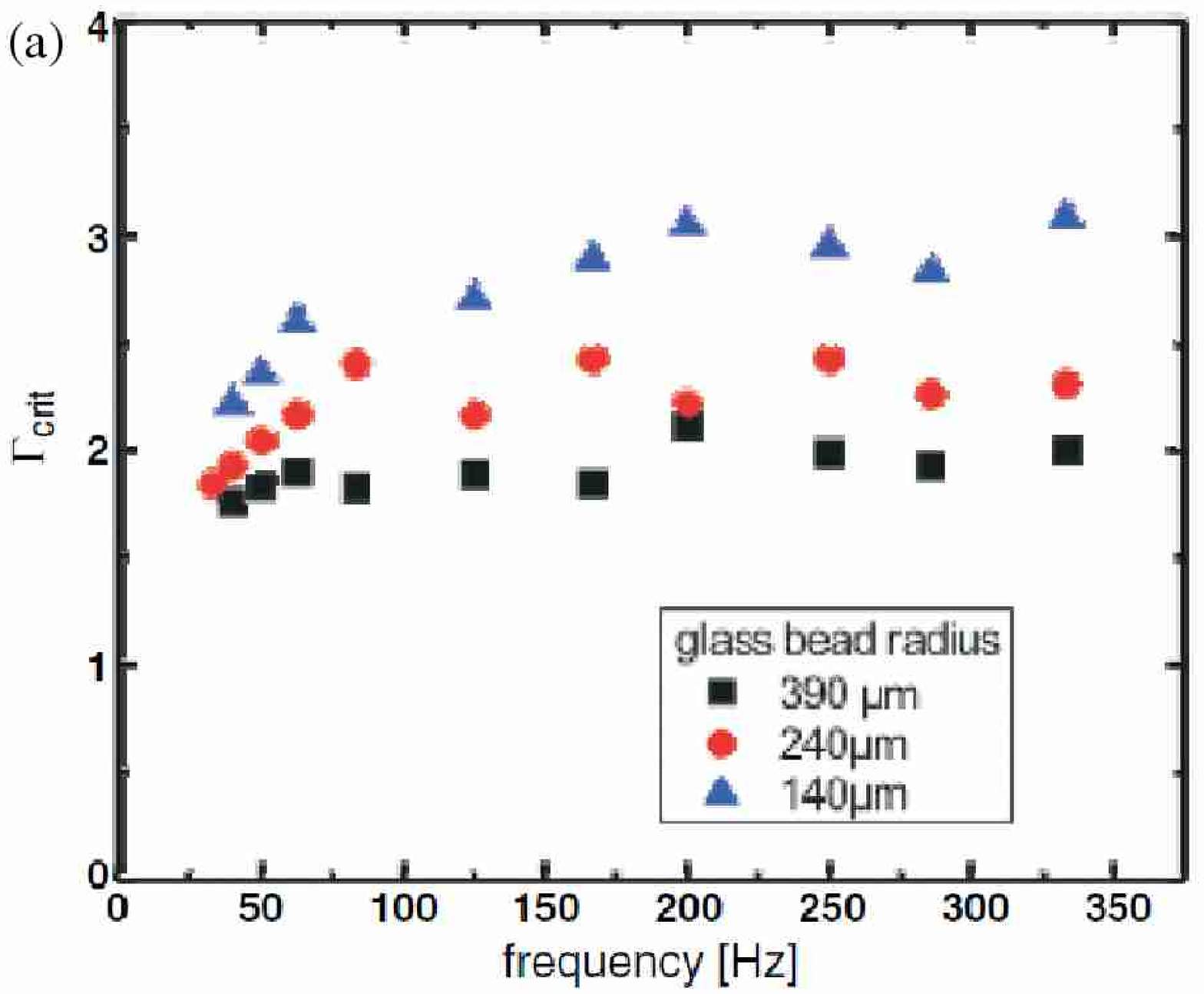}
\includegraphics[width=0.51\textwidth]{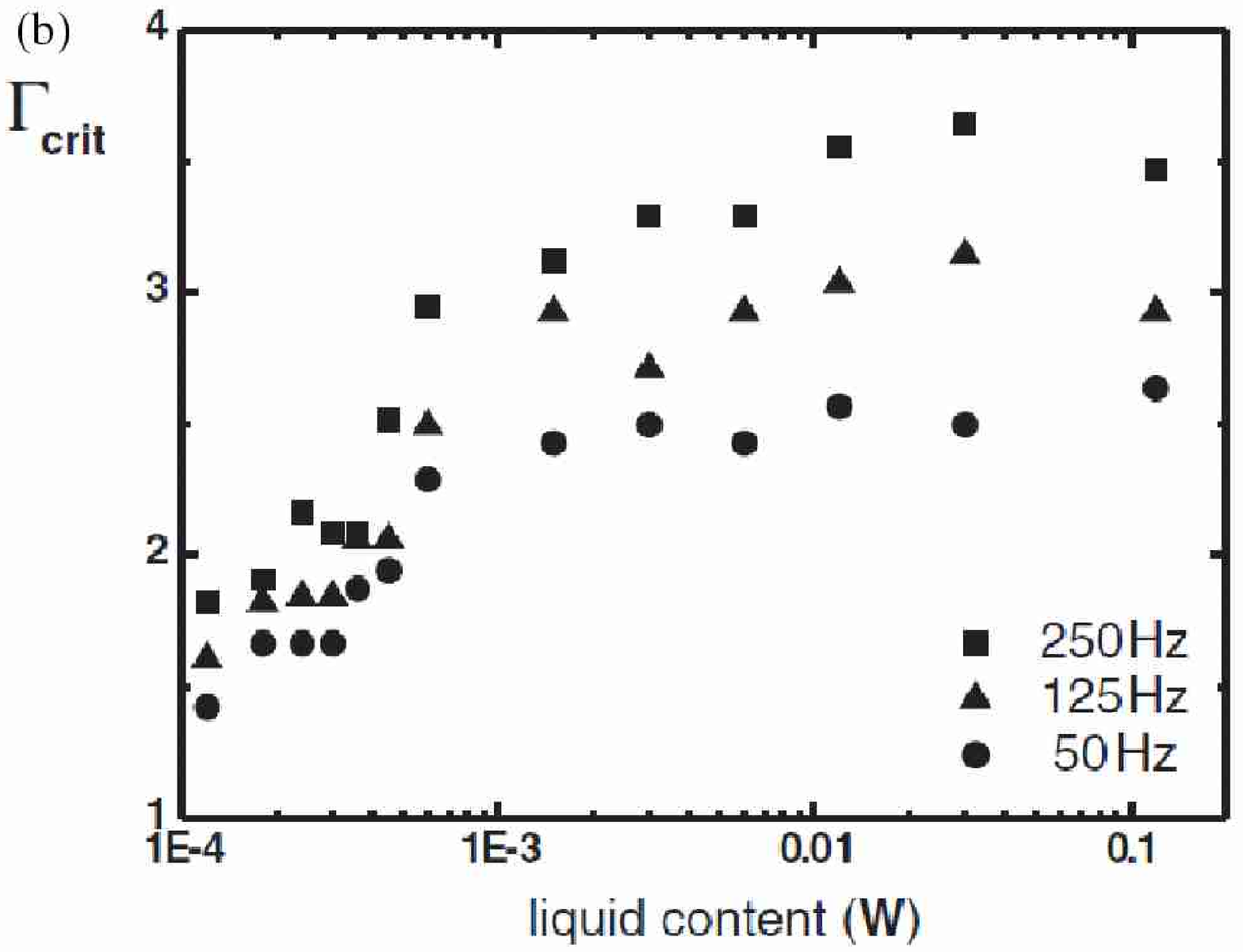}
\end{center}
\caption{
(a) Critical acceleration 
$\Gamma_{\rm crit}$
versus frequency $f$ for three different
bead radii, with water content $W=0.5\%$.
(b) $\Gamma_{\rm crit}$
versus liquid content $W$
for three different frequencies.
The particle radius R is 140 $\mu$m.
From \cite{Fournier05}.
}
\label{Fig:Schell04Fig1}
\end{figure}

For wet granular media, the cohesion force introduces 
an additional force or energy scale. 
It is non-trivial to determine
which parameters would better 
characterise the resulting behaviour.
Recently, 
experiments on this subject
have been conducted \cite{Schell04,Fournier05},
where fluidisation of glass beads wet by water
under vertical vibration has been investigated.
The critical dimensionless 
acceleration $\Gamma_{\rm crit}$ 
for fluidisation is found to 
depend on the frequency $f=\omega/2\pi$,
the particle radius $R$,
and the liquid content $W$
defined as the ratio of the liquid volume to the total volume,
as shown in Fig.~\ref{Fig:Schell04Fig1}.
Here, $W$ is defined as the ratio of 
the liquid volume to the total volume.
They \cite{Schell04,Fournier05} 
found that $\Gamma_{\rm crit}$ 
depends weakly on $f$, but
becomes constant for sufficiently high frequency. 
In this high frequency regime \cite{Schell04,Fournier05},
$\Gamma_{\rm crit}$ 
is smaller for larger $R$ for a given $W$ (Fig.~\ref{Fig:Schell04Fig1}(a)), 
while it increases
with $W$ for a given $R$
(Fig.~\ref{Fig:Schell04Fig1}(b)). 

Ref. \cite{Schell04,Fournier05} interpreted their 
results by considering the 
cohesion force due to liquid bridges
between grains
as well as between grains and the container wall.
The increase of $\Gamma_{\rm crit}$ 
with $W$ was found \cite{Schell04} to be proportional to
the increase of the cohesion force $F_{\rm bridge}$ per 
liquid bridge, which they estimated from $W$.
It is natural that $\Gamma_{\rm crit}$
increases with increasing $F_{\rm bridge}$,
but \cite{Schell04,Fournier05} found that the proportionality 
is not trivial: Another possibility is that 
$\Gamma_{\rm crit}$ is ruled by the energy
required to break the bridge, whose dependence on $W$
should be different from the dependence of the force. 
In the case of the dependence upon $R$, 
\cite{Schell04,Fournier05}
explained it by considering the shear stress 
due to the formation of liquid 
bridges between grains and side walls;
the bulk material tends to move together due to the cohesion,
but if the shear from the side wall is large enough,
the shear stress makes it possible to deform 
the bulk sample and to induce the fluidisation.
Following their interpretation, the 
decrease of $\Gamma_{\rm crit}$
upon increasing $R$ is because 
the force per bridge is proportional to $R$ but the 
number of liquid bridges per area is proportional to 
$1/R^2$, which gives the shear stress 
proportional to $1/R$.
They \cite{Schell04,Fournier05} 
also studied the fluidisation 
with the container whose walls are covered by the 
hydrophobic material so that liquid 
bridges cannot be formed with the side walls, 
and they found that the material cannot 
be fluidised up to $\Gamma=20$.

\paragraph{Segregation}
Segregation of grains by their size
is one of the 
most striking phenomena of granular materials.
The most well-known example is the Brazil-nuts effect, 
i.e., the large Brazil-nuts always appear on top 
of mixed-nuts containers. 
After two or more kinds of 
grains are mixed together,
size segregation occurs very easily
when they are excited by an external energy input
in, e.g., a vibrated container or chute flows. 
Size-segregation has been studied for dry granular material 
for various types of mixtures and excitations 
(e.g.,
\cite{Mobius01,Hong01,GranularSeg,Ottino00}).
In wet granular materials,
the cohesion tends to suppress the segregation
by sticking grains together,
but at the same time, various factors 
come into play in such a dynamical situation,
as we will see below.

The size-segregation 
in wet granular materials
that flow down an inclined plane
has been investigated in \cite{Samadani00,Samadani01}.
They used binary mixtures of glass beads with several 
sets of diameters. Also water, glycerol, 
and other kinds of liquid were used to observe the effect of 
the viscosity as well as the surface tension.
They found that segregation is basically 
suppressed by increasing the liquid content,
which is natural because the cohesion tends to
suppress the particles' relative motion.

Samadani and Kudrolli \cite{Samadani00,Samadani01}
investigated the phase diagrams of segregation
in the space of particle diameter ratio and
liquid content. 
These found a qualitative difference of phase diagrams
between water and glycerol:
the surface tensions are almost the same for 
these two kinds of liquid, but the 
viscosities are different, and  
the segregation is lower at higher viscosity.
This is because \cite{Samadani00,Samadani01}
viscosity tends to reduce the velocity fluctuations
required for segregation.

Samadani and Kudrolli \cite{Samadani00,Samadani01}
estimated the viscous force between 
moving particles connected by a liquid bridge 
by using the Reynolds lubrication theory, 
in which the hydrodynamic effect in 
thin space is taken into account
\cite{LubricationE,Tribology,Pitois00,Pitois01}.
There \cite{Samadani00,Samadani01}, 
the force is proportional to the relative velocity 
between moving particles, and they found that
the viscous force can be comparable with the cohesive force
for the characteristic velocity scale under gravity.
This also suggests that the viscosity is 
very important for dynamical behaviour like segregation.

\begin{figure}
\begin{minipage}{0.3\textwidth}
\includegraphics[width=\textwidth]{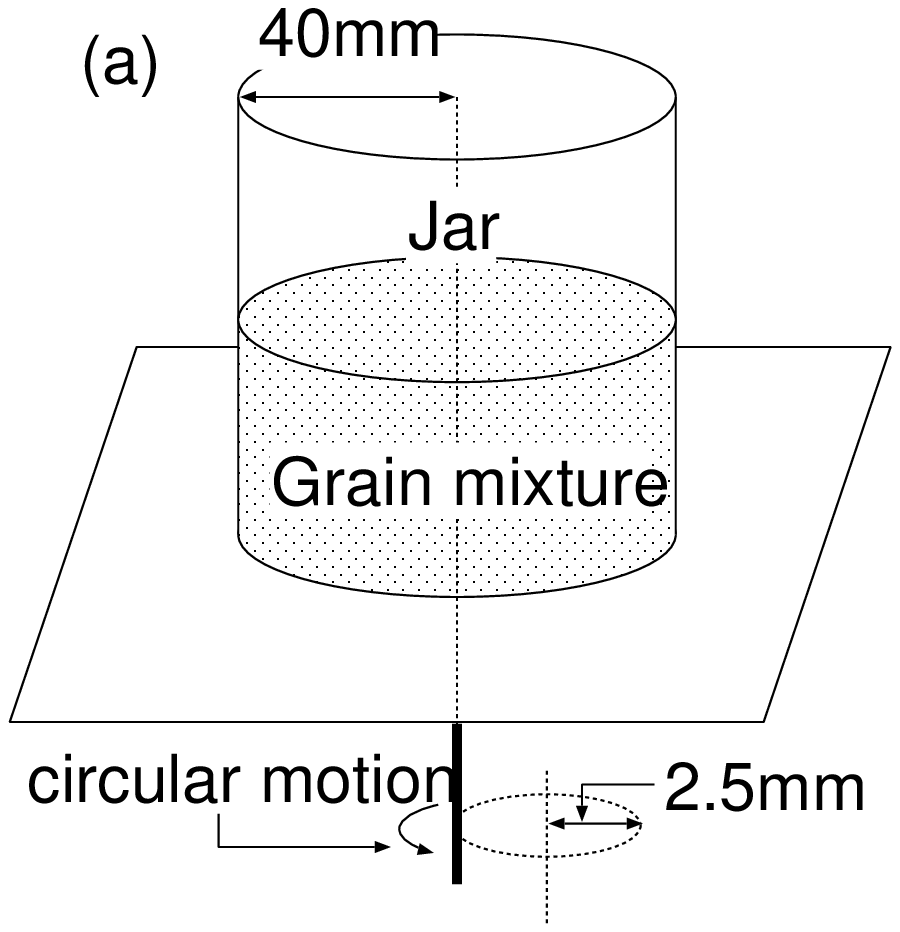}
\end{minipage}
\hfill
\begin{minipage}{0.6\textwidth}
\includegraphics[width=\textwidth]{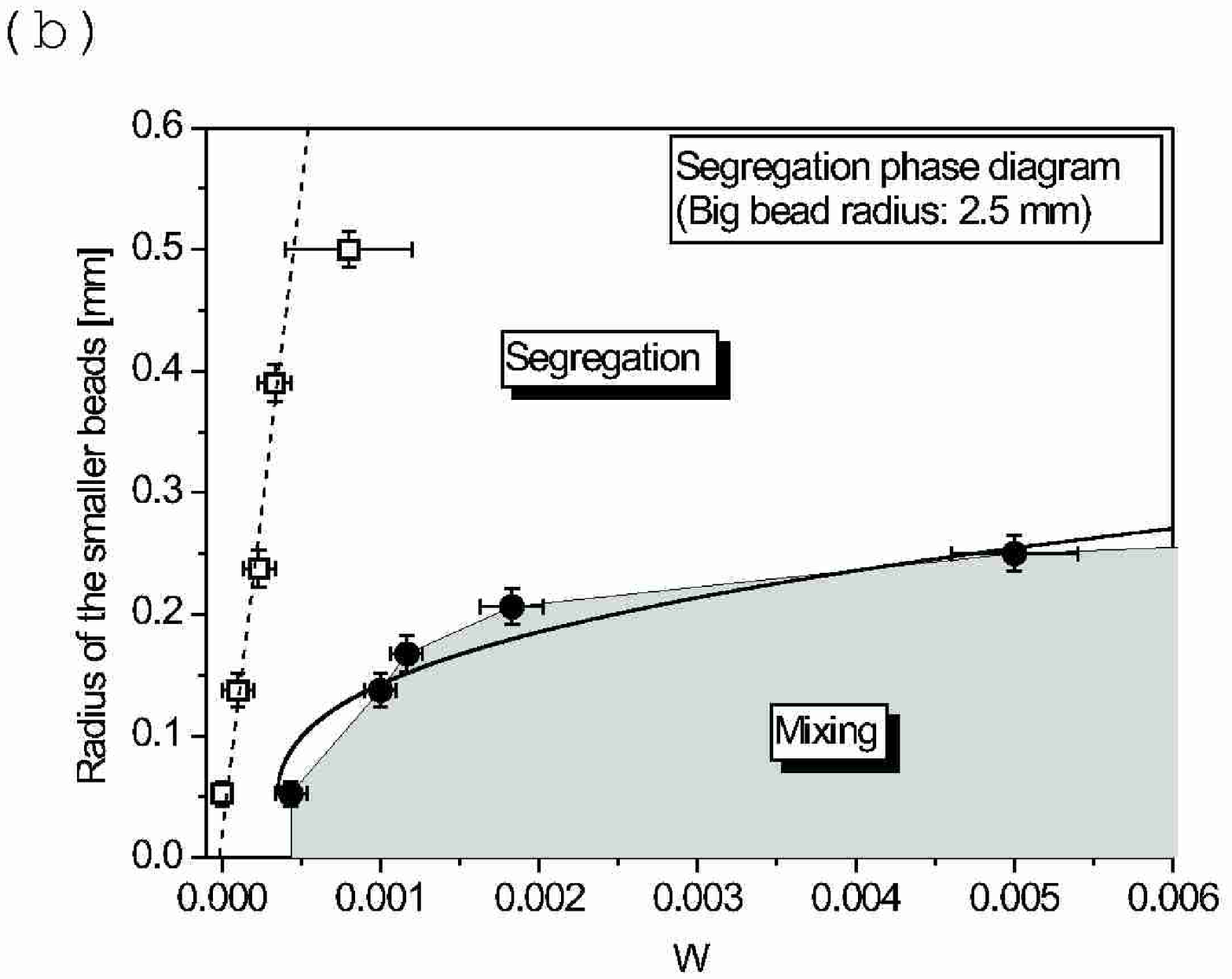}
\end{minipage}
\caption{
(a) Schematic description of the experimental setup
in \cite{Geromichalos03,Herminghaus05}.
The jar experience rapid, horizontal circular motions 
(20 revolutions per second),
where the centre of the jar follows a small circle
as shown in the figure
(The jar itself is not spinning around its centre).
(b) Particle-size segregation phase diagram
for binary mixtures of wet granular materials.
The radius of the bigger beads is fixed, and
the vertical axis shows the radius of the smaller beads.
The horizontal axis shows the amount of liquid 
via the dimensionless quantity $W$, 
defined as the volume of the liquid divided 
by the total volume of the small beads (including the space
between the beads).
The segregation does not occur in grey region.
The dashed steep line separates 
the region where the segregation is enhanced
by adding water (left) from the region 
where the segregation decreases or stays the same (right).
From 
\cite{Geromichalos03}. 
}
\label{Fig:Geromichalos03Fig3}
\end{figure}

On the other hand, Geromichalos {\it et al.} 
\cite{Geromichalos03,Herminghaus05} 
investigated size segregation in granular media
in a jar driven by a horizontal circular motion 
(Fig.~\ref{Fig:Geromichalos03Fig3}(a)).
They found that segregation sometimes occurs partially,
and they investigated the degree of segregation
by measuring the fraction of the
mixture zone (i.e., the region where no segregation occurs)
to the total sample.
From the experiments,
they distinguished three regimes:
(Fig.~\ref{Fig:Geromichalos03Fig3}(b));
the {\it gaseous} regime,
where an increase of the liquid content 
enhances size-segregation;
the {\it intermediate} regime,
where segregation occurs, but 
increasing the liquid content 
suppresses segregation;
and the {\it viscoplastic} regime,
where no segregation occurs.
By increasing the liquid content
with a fixed ratio of radii, 
the system state evolves
from gaseous, intermediate, to the viscoplastic regimes.

The behaviour in the gaseous regime 
found in \cite{Geromichalos03}, where 
segregation is enhanced 
by increasing liquid content,
seems to contradict the results in
\cite{Samadani00,Samadani01}.
Geromichalos {\it et al.} \cite{Geromichalos03} discussed that,
in the gaseous regime,
the energy dissipation is enhanced by the breakage 
of a number of tiny liquid bridges, 
while the cohesion is not enough to strongly affect the 
dynamics. 
The larger energy dissipation tends to 
enlarge the energy difference 
between large and small grains,
resulting in an enhanced segregation.
As liquid content is increased,
the liquid bridge becomes larger and the
kinetic energy becomes insufficient
to break the bridges \cite{Geromichalos03}. Then,
segregation is suppressed and
the system goes from the intermittent regime to 
the viscoplastic regime.

\subsubsection{Shear experiments for various liquid content}
\begin{figure}
\begin{minipage}{0.4\textwidth}
\includegraphics[width=\textwidth]{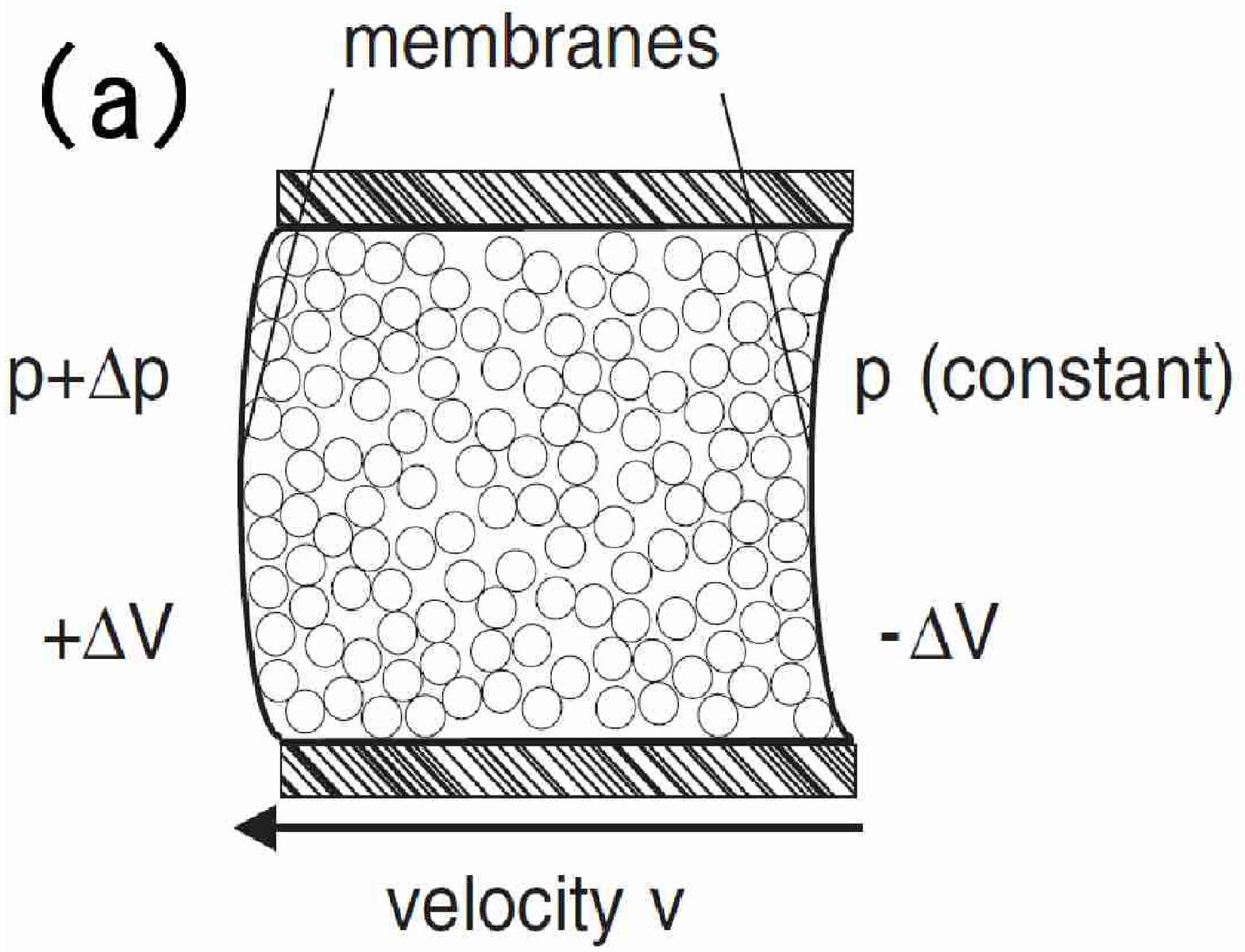}
\centerline{\includegraphics[width=0.5\textwidth]{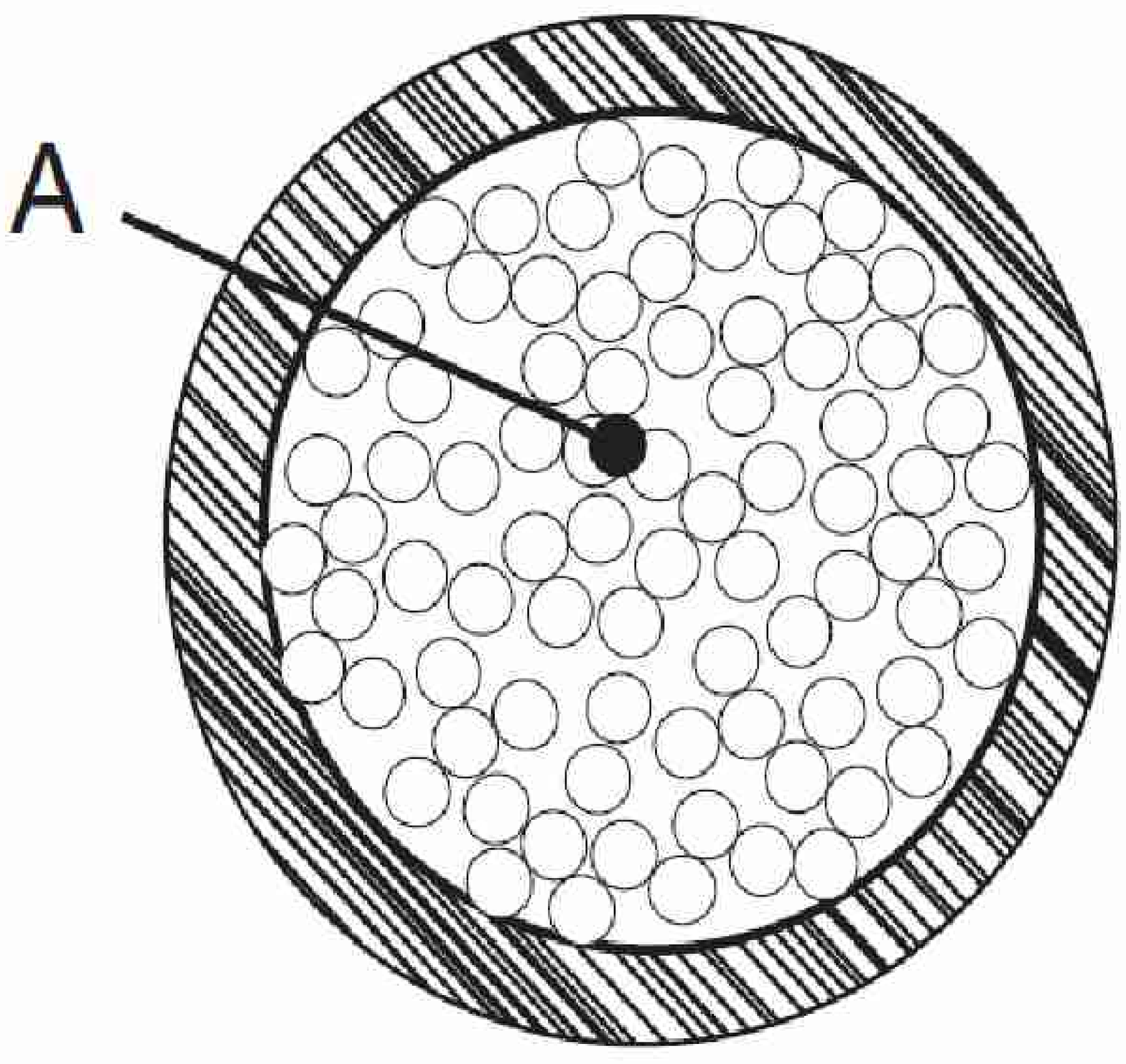}}
\end{minipage}
\hfill
\begin{minipage}{0.48\textwidth}
\includegraphics[width=\textwidth]{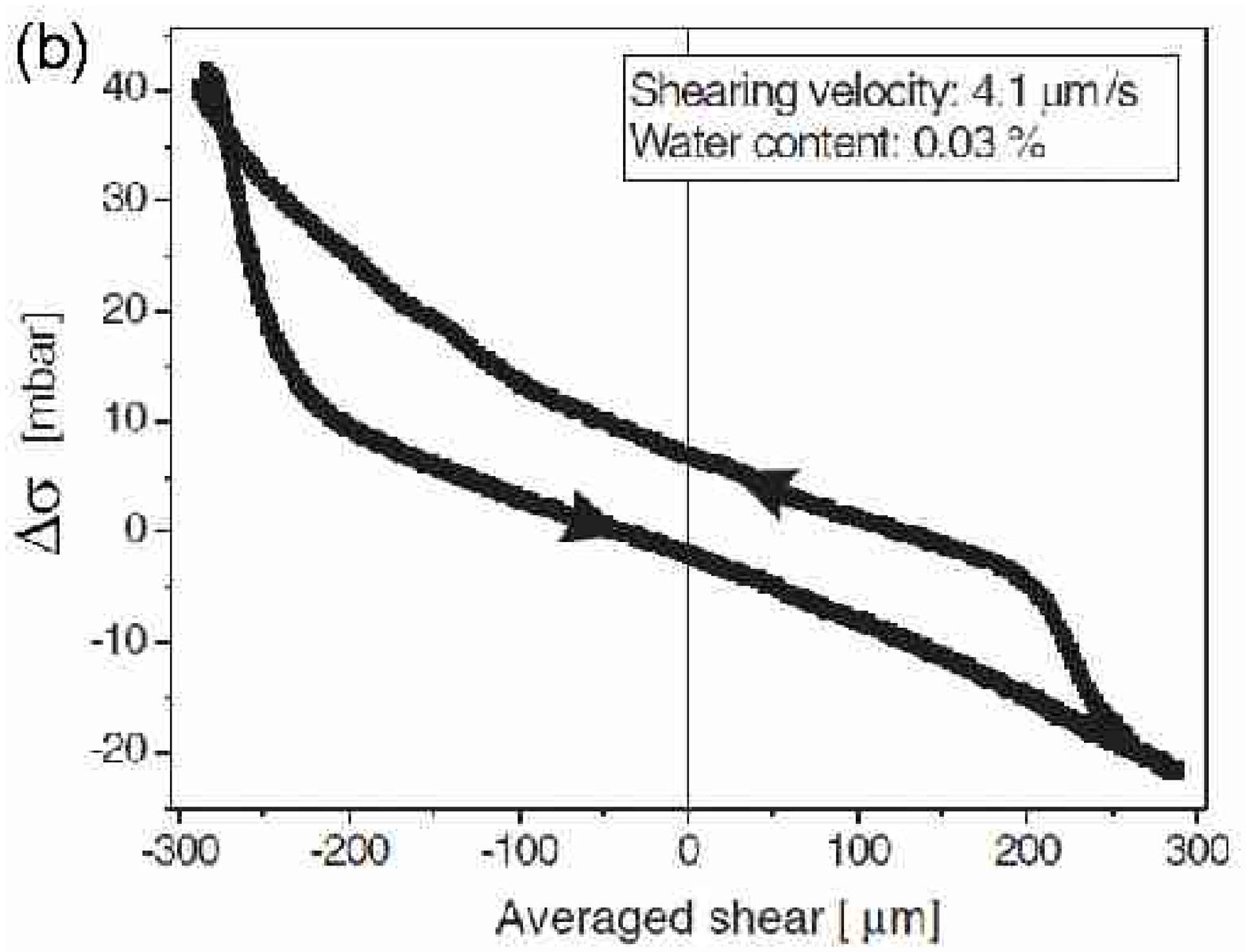}
\end{minipage}
\caption{(a) Experimental setup to shear 
wet granular materials in a cell,
showing a slice of the shearing cell seen from the side
and from the front.
(b) The hysteresis loop 
of the pressure difference $\Delta \sigma$
versus the average shear $\Delta V/A$
for densely packed glass beads.
From 
\cite{Fournier05}.
}
\label{Fournier05Fig7}
\end{figure}
\begin{figure}
\begin{center}
\includegraphics[width=0.6\textwidth]{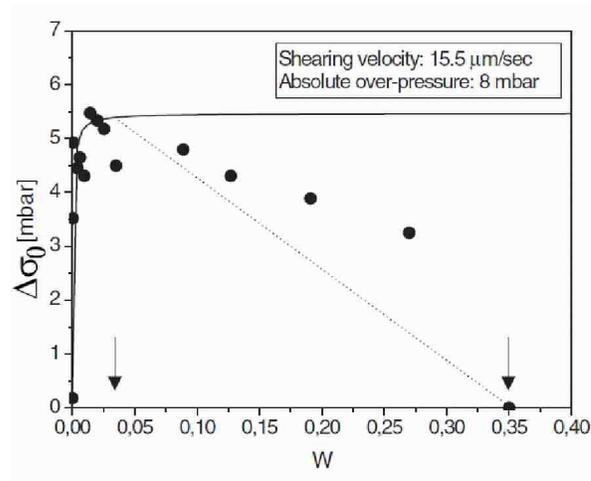}
\end{center}
\caption{
Dependence of $\Delta \sigma_0$
upon the amount of the liquid $W$.
From \cite{Fournier05}. 
}
\label{Fournier05Fig10}
\end{figure}
\begin{figure}[h]
\begin{center}
\includegraphics[width=0.5\textwidth]{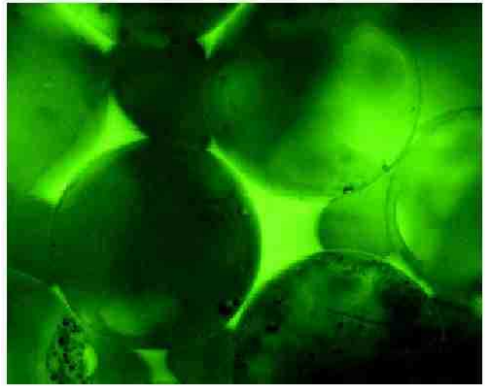}
\end{center}
\caption{
Fluorescence microscope images of liquid distribution 
between glass beads of diameter 375 $\mu$m for 
$W=0.03$, ``liquid clusters'' are formed.
From \cite{Fournier05}. 
}
\label{Fournier05Fig11}
\end{figure}

Recently, Fournier {\it et al.} \cite{Fournier05} investigated the 
response of wet granular materials under shear 
by using a unique experimental setup, shown in Fig.~\ref{Fournier05Fig7}.
A cylindrical cell was filled with
spherical glass beads wet by water, 
where the flat sides of the cell (area $A$)
each consist of a thin latex membrane.
Adjacent to each membrane was a cylindrical chamber 
filled with liquid, connected to a syringe. 
When the pistons of the syringe are moved at equal speed, 
the membranes are deformed by a volume
$\Delta V$ (Fig.~\ref{Fournier05Fig7}(a)
shows a lateral cut of the cylindrical cell (top) 
and a vertical cut (bottom)),
which allows them to shear the granular material under 
a constant volume at a controlled speed.
The pressure at each membrane is measured during the shear.
The pressure difference $\Delta \sigma$ between 
the two membranes 
versus average shear $\Delta V/A$ 
is shown in Fig.~\ref{Fournier05Fig7}(b)
for densely packed glass beads (packing fraction $\nu \approx 0.63$),
showing hysteresis. 
The vertical width of the hysteresis loop reflects the 
resistance to shear of the material.
They defined $\Delta \sigma_0$ as the
vertical width of the hysteresis loop 
at zero strain divided by two, and measured $\Delta \sigma_0$
for various liquid content and shear rate.
The dependence of $\Delta \sigma_0$ upon the amount of wetting 
liquid is shown in Fig.~\ref{Fournier05Fig10}. 
Following their parametrisation,
the horizontal axis shows $W$, the ratio of 
the liquid volume to the total volume.
For a given packing fraction $\nu$,
the liquid saturation $S$ is given by $S=W/(1-\nu)$, thus 
$W=0.35$ is almost in a 
saturated state ($S\approx 0.95$) when $\nu=0.63$.
We can see that $\Delta \sigma_0$ increases rapidly 
as $W$ is increased from zero, but it starts to drop 
at $W\approx 0.04 \ (S\approx 0.1)$ and goes to zero at $W\approx 0.35$.
They observed the liquid  
distribution for small liquid content
by index matching techniques, and
found that $W\approx0.03$ is 
the point where liquid bridges start to 
merge to form a liquid cluster
(Fig.~\ref{Fournier05Fig11});
beyond that point, the interfaces between 
the air and liquid start to decrease, 
which may cause the decrease of the cohesive force.

Reference \cite{Fournier05}  also investigated the shear rate dependence
of $\Delta \sigma_0$. Within the investigated 
range of the shear rate, $\Delta \sigma_0$ decreases 
for larger shear rate,
while the dry granular material 
does not show shear-rate dependence.
A possible origin of the shear-rate dependence 
is the time dependence of liquid motion as described in 
section \ref{Tegzes},
which affect the temporal evolution of liquid bridges.

In addition, they \cite{Fournier05} 
claim that the increase 
of the shear stress is not solely coming from 
the frictional effect, as in the Mohr-Coulomb picture described in section 
\ref{AngleofRepose}, 
but it can be understood by just 
considering the cohesion that arises from the liquid bridges.
One of the tests they conducted to confirm their claim
is to measure the absolute pressure dependence of $\Delta \sigma_0$;
if the frictional shear stress proportional to the
cohesive normal stress $\sigma_c$ is responsible for the 
fact that $\Delta \sigma_0$ is greater than that of the dry 
granular material, then applying a unidirectional pressure comparable
with $\sigma_c$ for {\it dry} granular material 
may cause a similar effect to increase $\Delta \sigma_0$.
However, they found that applying absolute pressure to the sample
is not enough to make $\Delta \sigma_0$ increase.
Though the fact that the applied 
pressure is unidirectional while
the cohesive pressure is uniform
may be one of the causes of this difference \cite{Herminghaus05}, 
their results indicates that
the origin of the higher yield shear stress
in wet granular materials need to be investigated further.

The response of materials against shear is 
a fundamental property
to understand the material's rheology.
Further experiments of sheared wet granular material
in various setups would be useful.

\subsubsection{
Dynamics of  
wet granular media: practical applications}
There are many research fields 
that investigate the dynamics
of wet granular media from a practical point of view.
Below we summarise a few examples.

\paragraph{Agglomeration processing:
grains binded by liquid} \label{Granulation}
During an agglomeration or granulation process,
particles lump or agglomerate together into
larger, semi-permanent aggregates called granules
\cite{GranulationReview,GranulationProceedings,Cates04}.
There are several methods to make granules.
One of them is 
to spray a binder liquid onto dry powder 
and mix them in, e.g., a tumbling drum
so that clumps of particles binded by liquid grows.
The granules may either grow or break down
when they collide.
Another method is to strongly shear 
a very dense mixture of binder liquid and powder (dense paste), 
and then air comes into the paste to form lumps.
In either method, the main source of the cohesive force
that binds powder particles together is the capillary force.

Numerous dynamical experiments have been performed
on granulation processing:
The dynamics of liquid distribution during granulation, 
the growth of granules, 
the shear rate dependence of the growth rate, 
the collision velocity dependence of breakage,
and so on.
Though most of the experimental setups seem
rather complicated, 
the accumulated knowledge provides
many insights on the dynamical properties of wet granular media.
However, it is beyond the scope of this brief review
to summarise all experiments 
in this subfield.
Interested readers 
should consult a rather recent review on this topic
\cite{GranulationReview},
a conference proceedings \cite{GranulationProceedings},
and references therein.

In addition, the properties of sheared dense paste 
are important not only 
in the granulation process, but also in, e.g., ceramics engineering,
where the rheology of paste has been investigated.
The importance of cohesion due to the capillary 
effect in the rheology of paste
has often been pointed out 
(e.g., \cite{CeramicProcessBook,VanDamme02}),
and the knowledge about its rheology also provides 
insights about wet granular media.

\paragraph{Geological events}
Geology is one of the largest research fields 
on wet granular media.
The failure criterion of soils
and also their dynamical behaviours 
are very important to understand.
Significant, and sometimes catastrophic and tragic, 
examples include land slides, debris flows, 
and liquefaction of ground 
(see, e.g., \cite{Iverson97,Dikau,Coussot,
SoilJapaneseBook,Ishihara93}).

Most of the past work on 
debris flows focused on the flow of
soils saturated by water.
In studies of debris flows,
knowledge about dry granular media
has been incorporated into the analysis, and at the same time, 
the significant effects of the liquid lubrication and
the liquid viscosity on debris flows have been investigated
\cite{Iverson97}.
Liquefaction of wet soils triggered by earthquakes
is also often studied in saturated situations \cite{SoilJapaneseBook}.
Systematic studies of wet granular media
from partially-wet to the completely wet state
should also provide useful insights to understand 
these important geological events.

\section{Summary and open questions}
\subsection{Effect of the liquid content on 
quasistatic behaviour}
\label{SummaryCohesion}
\begin{figure}
\begin{center}
\includegraphics[width=0.6\textwidth]{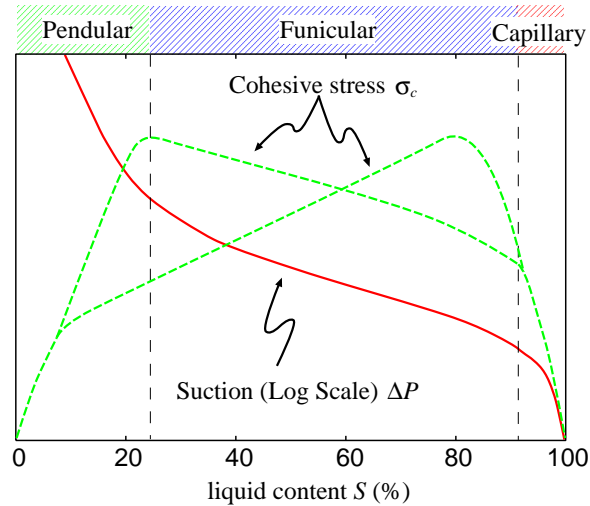}
\end{center}
\caption{
Schematic diagram summarising the variations of
the suction $\Delta P$ and 
the cohesive stress $\sigma_c$, 
upon changing the liquid content $S$.
}
\label{Fig:Schematic}
\end{figure}

The cohesive force due to the presence of a
liquid arises in the pendular,
funicular, and capillary states.
This cohesion gives rise to a finite cohesive stress
in quasistatic experiments.
Figure \ref{Fig:Schematic} shows a schematic 
diagram summarising how the suction $\Delta P$
and the cohesive stress $\sigma_c$
vary upon liquid content.

The dependence 
of the suction $\Delta P$
on the liquid content $S$
is given by a red solid line, and
its slope changes significantly near the phase boundaries between 
the pendular, funicular, and capillary states,
as discussed in subsection \ref{phases}.

The cohesive stress $\sigma_c$ versus $S$
is schematically shown by the green dashed lines.
The overlapping dashed curves for low and high $S$ are better established,
while the non-overlapping intermediate curves 
vary significantly between experiments.
The cohesive stress $\sigma_c$ increases 
as we add more and more liquid to initially dry grains,
shown by the positive 
slope, for small $S$, of the $\sigma_c$ line 
in the pendular state.
At the opposite end, cohesion becomes zero 
for completely saturated granular media ($S=100\%$),
schematically shown by the green dashed line with negative slope 
for $S$ close to 100\% in the capillary state.

In the funicular state,
the cohesive stress $\sigma_c$ dependence
on the liquid content $S$
is not clearly understood.
The cohesive stress $\sigma_c$ may either increase or decrease with $S$
in the funicular state. 
These possible curves are shown by the green dashed lines,
where both lines connect to single lines 
in the limits when $S\to 0$ and $S\to 100\%$.
We see that there would be at least one maximum of 
the cohesive stress at a certain liquid content, though
it is not clear, a priori, the location of the peak.

\subsection{Open problems}
There are a number of open problems
on mechanical properties of wet granular media,
some of which have already been mentioned in the text.
In this subsection, we list a few examples of 
open problems, in order to 
encourage future studies in this area.
\subsubsection{Jamming}
\begin{figure}
\begin{center}
\includegraphics[width=0.7\textwidth]{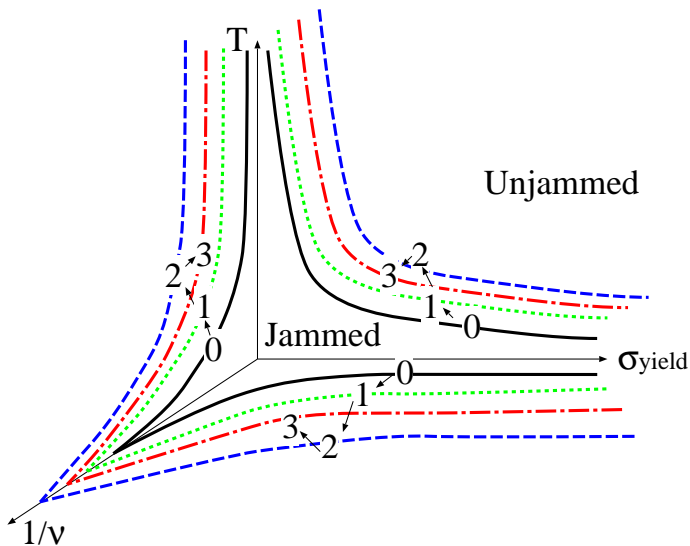}
\end{center}
\caption{
Schematic diagram of 
a possible jamming phase diagram 
of granular media for various liquid contents.
Number 0 (black solid line): Dry . 
Number 1 (green dotted line): Small amount of liquid (slightly cohesive).
Number 2 (blue dashed line): ``Optimal'' amount of liquid (the most
 cohesive,
and the largest phase-space volume of the jammed phase).
Number 3 (red dash-dotted line): Large amount of liquid 
(less cohesive than the optimal case).
}
\label{Jamming}
\end{figure}

As summarised in subsection \ref{SummaryCohesion},
in the quasistatic regime, the cohesion 
in wet granular media 
has a maximum for a certain liquid content,
though whether the maximum is located
at a rather small or large liquid content
is not known a priori.
This variable cohesion will 
affect the ``jamming'' properties of the granular assembly.
Thus, an interesting problem
worth considering 
is the jamming phase diagram in wet granular media.

A jammed state occurs when
a disordered system subject to an 
external force is caught in a
small region of phase space 
with no possibility of escape 
\cite{Liu98,JammingBook,DAnna01,DAnna03,OHern02,OHern03,
Trappe01,Valverde04}.
In order to unjam a jammed material,
a finite yield stress or fluctuation energy is needed,
which forces the elements to escape from 
the phase space region where they are trapped.
Granular media at rest is 
a typical jammed material, where we need 
a finite yield stress or external energy input
in order to let the grains flow or move.
There are many kinds of materials that show
jamming, such as dense colloids, 
and pinned vortices in superconductors.
Jamming also plays a role in 
the mechanical properties of wet granular media.
For example, Cates {\it et al.} \cite{Cates04} recently
tried to understand the physics of granulation (section \ref{Granulation})
in a highly-sheared dense paste in terms of a jamming 
transition under shear.

Liu {\it et al.} \cite{Liu98} 
proposed a very schematic phase diagram for jamming,
in the space of packing fraction $\nu$ (density), 
external load, and temperature
or fluctuation energy. The purpose of this 
schematic phase diagram is to 
try to summarise a unified view of the jamming phenomena,
observed in various media.
A schematic diagram of our jamming phase diagram
is shown in Fig.~\ref{Jamming};
the solid black curves, 
labelled with the number 0, was proposed 
by \cite{OHern02,OHern03} for non-cohesive (dry) granular media.
The axes shown are:
inverse packing fraction $1/\nu$,
the yield stress $\sigma_{\rm yield}$,
and the fluctuating energy $T$
from an external energy source, such as vibration.
One of the features of jamming in dry granular media
is that the behaviour in the $T=0$ plane, 
i.e., no external energy, can be easily 
investigated. The ``jammed'' phase-boundary
intersects the inverse packing fraction $1/\nu$ axis at zero yield stress
and temperature,
and the packing fraction at the intersection
corresponds to the random closed-packing density \cite{OHern02,OHern03}.

Obviously, attractive interactions modify the 
jamming properties.
Trappe {\it et al.} \cite{Trappe01} investigated jamming in colloidal systems
with varying attractive interaction potentials;
they found that the jammed phase boundary 
shifts to higher temperatures for stronger attractive potentials,
because the attractive potential characterises
the energy that particles need to escape from the 
phase space where they are trapped.
Jamming in cohesive powders was investigated by
\cite{Valverde04}; the powder particles were large enough 
so that thermal noise was negligible.
It was found that the critical jamming density 
for zero fluctuation energy and yield stress
for cohesive powder is smaller than the critical
density in non-cohesive granular media.
This is because the cohesion 
made it possible to form
very sparse but stable 
clusters of powder.
Indeed, a percolating cluster appears
in their system for relatively low densities.

Considering these results, 
the cohesion present in wet granular media 
will decrease the density needed to jam the system.
To break a cluster in 
wet granular media, a large enough 
external fluctuation energy is needed,
and this energy will increase with cohesion.
The yield stress will increase with cohesion 
as described by the Mohr-Coulomb criterion,
where the yield shear stress increases
with the cohesive stress $\sigma_c$.
These considerations imply that, 
for cohesive granular media,
the volume of phase space occupied by the jammed
state expands as the cohesion increases.

How is the jammed phase diagram
for granular media modified by adding liquid?
A schematic phase diagram is shown in Fig.~\ref{Jamming}.
The number in each curve there shows how the behaviour changes
as we increase liquid content.
When a small amount of liquid is added, 
the phase boundaries are slightly shifted 
so that the jammed phase expands 
in all three directions of $1/\nu$, $\tau$, and $T$
(number 1, green dotted lines).
As liquid is further added, the cohesion {\it increases},
and the boundary will shift further.
At a certain liquid content, the
cohesion will reach a maximum, 
corresponding to the {\it largest jammed region}
in phase space (number 2, blue dashed lines).
As we further increase the liquid content,
the cohesion {\it decreases}, and the 
phase-space volume of the jammed region decreases
(number 3, red dash-dotted line).
Namely, as far as cohesion is concerned, 
there is an optimal liquid 
content that maximises the jammed region.
This argument
does not consider the effects of liquid
other than cohesion.
For large enough liquid content, 
the lubrication effect may affect this conclusion,
and could drastically shrink the size of the jammed region.
The liquid content dependence of the jamming behaviour
in wet granular media
is an important and interesting 
phenomenon that has not been systematically investigated
and constitutes a fertile new direction of research. 

\subsubsection{Statistical mechanics approach}
Recently, many physicists are trying to 
describe dry granular systems using statistical mechanics approaches,
especially in the jammed regime (For a recent review,
see \cite{Richard05}). 

Edwards \cite{Edwards,Edwards02} postulated that 
a dense granular assembly 
under small external perturbations can
take all possible jammed configurations,
and the density can be described by suitable ensemble averages over 
its blocked states. Later, Nowak {\it et al.} \cite{Nowak98} investigated
the compaction of slowly tapped granular assemblies,
and found there exists a ``reversible regime'': 
When a loosely packed configuration is tapped,
large voids are removed, which results in irreversible
motion of grains (the initial loosely packed configuration
cannot be reproduced by further tapping).
Once the memory of this initial configuration is
lost, after large enough tapping,
the density is determined by the ratio $\Gamma$
of the tapping amplitude to gravity,
which is in the reversible regime; 
in this regime, the density is lower for larger $\Gamma$.
The existence of a reversible regime is non-trivial
in granular matter, where the frictional force depends on history.
The Edwards' scenario only considers non-history dependent
situations, and the existence of such a regime gives hope
that this scenario would work in the reversible regime.
Research to test this theory has been done
\cite{Brey00,Tarjus04,Nicodemi99,Coniglio01,Makse02,Fierro02,Nicodemi02}, 
and there remains many open questions.

Another example of a statistical mechanics approach
is the recent experiment by \cite{DAnna03}, 
who observed the motion of 
a torsion oscillator immersed in a granular medium
perturbed by external vertical tapping,
and measured  the susceptibility
and the auto-correlation function of the motion 
of the oscillator.
For the equilibrium systems,
these two quantities are related by
the fluctuation-dissipation theorem and 
the temperature can be consistently determined.
They \cite{DAnna03} examined whether the same formalism works in 
the granular state, and found that
an ``effective temperature'' can be defined
from the susceptibility and the auto-correlation function.
Though this dissipative system is far from equilibrium, 
the analogy to the fluctuation-dissipation theorem 
in the experiment is remarkable. It is an interesting question 
whether such a relation 
can be extended to other systems or regimes
(e.g., wet grains).

In the case of wet granular media, the available phase space 
is larger and,
due to the cohesion, the history dependence is stronger than the dry case 
as mentioned in Section \ref{compaction} 
and Table \ref{Table:DryWet}. 
Future experiments probing the statistical mechanics 
in wet granular assemblies will 
provide more knowledge of its applicability and generality.
\subsubsection{Arches and contact-force fluctuations}
Arches (i.e., effective long-range interactions)
and the resulting large fluctuation of contact forces
have been extensively studied in 
several types of dry granular media 
(see, e.g., \cite{Hartley03,Albert00,Liu95,Howell99,Miller96,
Coppersmith96,Blair01}).
However, no such systematic studies exist
for wet granular media.
The clogging of cohesive powders 
in a hopper is often said to be an arching effect
\cite{Nedderman},
but it is rather rare to investigate
the arches directly in partially wet granular media.
This would be interesting 
because the change of the inter-particle forces
with liquid content
likely affects the strength and 
topology of the so-called 
``force chains'' or stress-line-networks
of the granular assembly.

\subsubsection{Simple experimental setups to 
study the dynamics of wet granular media}
Several experimental setups have been
widely performed to study the fundamental 
dynamics of dry granular media.
The widely-used setups include
vibrated granular media (e.g., 
\cite{Losert99,Melo95,
DAnna03,
Umbanhowar96,Pak94,Laroche89,DAnna01,
GranularSeg})),
rotating drums (e.g., 
\cite{Rajchenbach90,Rajchenbach02,Bonamy02,RajchenbachReview,
Benza93,Ristow96,Khakhar97,Baumann95}),
shear (e.g., \cite{Xu03,Mueth00,Miller96,Nasuno97}), and
inclined chutes (e.g.,
\cite{Poschel93,Mitarai02,Mitarai03,Mitarai04,
Mitarai05,
Pouliquen99,
Daerr99,Savage89,Komatsu01,Bretz92}).
These setups have played important roles in 
studying the fundamental dynamics and 
rheology of dry granular materials. 

Some of the setups have been used in recent experiments
of wet granular media, as discussed in 
section \ref{dynamicalbehaviours},
but not so many systematic studies have been done
so far. 
One of the difficulties in fundamental studies 
of wet granular materials is their strong 
tendency to be inhomogeneous.
It becomes difficult to induce particle motions,
due to cohesive force. 
In other words, the regions that were mobile
in the dry case become localised when wet,
and the bulk material becomes solid.
Similar behaviours are known for dry granular materials as
well (e.g., shear bands), but the localisation would be 
stronger for the wet case.
In addition, the distribution of liquid can 
also be inhomogeneous, 
especially for larger liquid content.
Nonetheless,
further studies on the dynamics of wet granular materials
(including their inhomogeneity)
using these rather simple setups
will certainly contribute to granular science.

\subsubsection{Numerical simulations}

The recent remarkable progress in
the study of dry granular material is 
partially 
due to numerical simulations.
Molecular dynamics simulations of soft-sphere 
or inelastic hard-sphere models help
clarify phenomena found in experiments
and get data in ``ideal'' situations.
In the case of wet granular media, however, 
its numerical simulation model 
for the wide range of liquid content
has not been established yet because of its complexity.

Some models 
of wet grains in the pendular state have been proposed 
for molecular dynamics simulations 
(e.g., \cite{Groger03,Schulz03,Lian98,Nase01,Hakuno93,
GranulationReview,GranulationProceedings}),
where most of them are based on 
soft-particle models for dry granular materials 
with elastic force and dissipation \cite{Cundall79}.
In models for wet grains, the effect of the liquid 
is added by assuming that liquid bridges 
are formed when grains are in contact, 
and cohesion and dissipation 
due to the liquid bridges are taken into account.
These types of models would be applicable to some extent
to situations where the amount of liquid is small
and the liquid mostly sticks to the grain surface.
However, as the liquid content is increased,
the liquid-bridge picture becomes invalid 
and liquid motion becomes relevant.
No available simple models has been yet proposed 
for this regime, as far as we know.

Even for dry granular media, 
the real granular particles are more complicated than
the ones used in
most computational models; still, the simple models have been
found to be very useful.
It is apparent that good models are also
needed for wet granular
 material.
The required level of realism of the model largely depends on 
the phenomena that one would like to understand.
To help the modelling,
more systematic experiments in simple situations
are necessary.

\subsubsection{Mechanical properties of snow}
The mechanical properties of snow have been
investigated for a long time
(e.g., \cite{Fukue76,Nicot04,DAnna00,Bartelt01,snowski,Shapiro97}).
It is important to understand snow properties
to reduce disasters caused by snow avalanches;
thus knowledge should also be
useful to design equipment
and constructions in snowbound areas \cite{Shapiro97}.

Snow is a type of granular material of grains of ice. 
Snow partially melts in many situations,
and pores are filled with both air and water;
it is a ``wet granular material''. 
The knowledge obtained about the dynamics of dry granular media
has been found to be useful to understand snow dynamics;
for example, the mechanism of size segregation 
is important to understand how to save skiers caught in an avalanche
\cite{Bartelt01}.  In terms of cohesion among grains,
wet snow avalanches has some aspects 
in common with avalanches in partially wet granular media
\cite{Tegzes03}. More research on wet granular media
should be useful to better understand the mechanical properties of snow.

However, there is a big difference between 
wet sands and snow; in snow, solid ice-bonds 
exist between ice grains. 
In new or low-density snows, ice grains 
slide over each other, while in old or dense snow,
the deformation of solid ice-bonds dominate the interactions
\cite{Bartelt01}. The formation, strength, and breakage of ice-bonds 
depend on many factors
such as the heat flux and motion of water or water vapours.
Because of this complex nature of interactions,
there remains many open questions in snow physics. 
For example, it is known that the density of snow
is not a good parameter to characterise snow properties,
and one needs to classify snow by considering
its microstructure \cite{Shapiro97}.
It will be useful to find a parameter 
that is easily measured and
can better characterise its mechanical properties.

\section{Conclusion}
Wet granular media have properties 
which are significantly different from
dry grains. These include enhanced cohesion
among grains leading to very 
steep angles of repose.
Wet granular media are
pervasive everywhere, 
including numerous industrial applications
and geological phenomena.
In spite of the ubiquitous 
presence of wet granular media and its importance,
relatively little is known about it.
This brief overview sketched some
of the physical properties of wet granular media, 
and identified several open problems for future studies.

\section*{Acknowledgements}
NM thanks H. Nakanishi for helpful discussions.
Authors thank G. D'Anna for his comments on an early version 
of the manuscript.
Part of this work has been done 
when NM was supported by the Special Postdoctoral Researcher 
Program by RIKEN.
NM is supported in part by a Grant-in-Aid for
Young Scientists(B) 17740262 from
The Ministry of Education, Culture, Sports,
Science and Technology (MEXT) and 
Grant-in-Aid for Scientific Research (C) 16540344 from 
Japan Society for the Promotion of Science (JSPS).
FN is supported in part by the National Security Agency (NSA)
and Advanced Research and Development Activity (ARDA) under Air
Force Office of Research (AFOSR) contract number F49620-02-1-0334,
and by the National Science Foundation grant No. EIA-0130383.

\end{document}